\documentclass[12pt]{article}

\usepackage{amsmath,amsfonts,latexsym,amssymb,amscd}
\usepackage{pslatex}
\usepackage[latin1]{inputenc}
\usepackage[T1]{fontenc}
\usepackage{graphicx} 
\usepackage{pspicture}
\usepackage{verbatim,amsthm,curves,graphics}
\usepackage{mathrsfs}
\usepackage{xcolor}

\usepackage{graphicx}

\begin{document}

\title{No-hair theorems in General Relativity and scalar-tensor theories}

\author{
 Stoytcho S. Yazadjiev$^{1,2}$\thanks{\tt yazad@phys.uni-sofia.bg} \; Daniela D. Doneva$^{3}$\thanks{\tt daniela.doneva@uni-tuebingen.de},
\\ \\
	{\it ${}^1$ Department of Theoretical Physics, Sofia
	University "St. Kliment Ohridski".} \\
{\it 5 J. Bourchier Blvd., Sofia 1164, Bulgaria} \\
{ \it ${}^{2}$ Institute of Mathematics and Informatics, 	Bulgarian Academy of Sciences, }\\	{\it Acad. G. Bonchev St. 8, Sofia 1113, Bulgaria}\\
{\it ${}^3$ Theoretical Astrophysics, Eberhard Karls University of T\"ubingen} \\
{\it 72076 T\"ubingen, Germany}}

\date{}

\maketitle

\begin{abstract}
In the present review, we consider the status of the classification of the vacuum, stationary and asymptotically flat black holes in scalar-tensor gravity. Contrary to the similar problem in general relativity, the black hole classification in scalar-tensor theories is much more challenging due to the very complicated character of the field equations and the very complex mathematical structure of the scalar-tensor gravity as a whole. We review most of the known no-hair results, and where possible new ones, as well demonstrate some of the difficulties that appear in our attempts to classify the black holes within scalar-tensor gravity. 
The proofs of the theorems and the underlying mathematical techniques are given in sufficient detail. To make the review self-contained we also present the vacuum black hole uniqueness theorems in general relativity and their proofs. 
\end{abstract}


\section{Introduction}

The classification of the stationary and asymptotically flat black hole solutions is a classical problem in general relativity (GR) and has a long history \cite{Robinson_2009, Mazur:2000pn}. The physical formulation of the problem is related to Wheeler's famous phrase ``black holes have no hair'' \cite{Ruffini:1971bza}. In other words, whatever matter originates the black hole, all information about it, disappears, except for a small set of asymptotically measurable quantities (charges). 
This physical formulation of the problem, known as the no-hair conjecture in GR, is however too broad and vague to be treated mathematically. As in most cases in physics, to have a mathematically more precise formulation of a given problem, we have to consider it in a more or less idealized framework. The natural idealization of the problem is to consider single black holes in vacuum (or electrovacuum). Under such simplification, the non-hair conjecture states that all stationary and asymptotically flat black hole solutions to the vacuum (or electrovacuum) Einstein field equations are fully specified by their conserved asymptotic charges -- the mass and the angular momentum (and the electric charge). In this formulation, the no-hair conjecture was rigorously proven and these 
are the famous uniqueness theorems \cite{Mazur:2000pn},\cite{Israel:1967wq} - \cite{Bunting_1983} (see also \cite{Heusler_1996}). The conjecture however fails in more general settings, i.e. in the presence of various matter fields, which is a well-known fact (see for example \cite{Droz:1991cx} - \cite{Hartmann:2001ic}). It is also interesting and worth mentioning that the no-hair conjecture even in its refined mathematical form is violated in higher dimensions, too. For example,  the stationary vacuum and asymptotically flat black hole solutions to the five-dimensional Einstein equation are no more fully specified in terms of their asymptotic charges (the mass and the angular momenta). A new datum called ``interval structure'' has to be added to the angular momenta in order to fully specify the black holes solutions and we refer the reader to \cite{Hollands:2007aj} and \cite{Nedkova:2024oqh} for more details. Nevertheless, the no-hair conjecture played a very important role in GR and the proven black hole uniqueness theorems are among the key results in modern GR.

In the last two decades, the so-called scalar-tensor theories of gravity attracted a lot of interest. They provide a rich framework for exploring gravitational physics beyond GR. In addition to the spacetime metric, the scalar-tensor gravity introduces one or more dynamical scalar fields which are coupled to various curvature invariants and/or to the matter fields. Black hole solutions within the framework of scalar-tensor gravity have also been intensively studied. In this way, the problem for
the classification of the possible stationary and asymptotically flat scalar-tensor black hole solutions arises naturally. The lessons learned from the pure general relativistic case suggest that we should first focus on the vacuum sector of the scalar-tensor theories if we want to make some progress in classifying the scalar-tensor black holes. In general, however, the black hole classification problem within the scalar-tensor gravity is much more difficult than in GR and poses formidable mathematical challenges. The difficulties are due to the very complicated character of the equations and the very complex mathematical structure of the scalar-tensor gravity as a whole. Even the vacuum sector of scalar-tensor theories is rather complicated
as it can roughly be considered to be the Einstein equations coupled to scalar fields which in turn can have various couplings to spacetime curvature.

The posing of the black hole classification problem in scalar-tensor gravity is highly non-trivial. For example, the scalar degrees of freedom are not obligated in general to share the symmetries of the spacetime metric. In other words, even when the spacetime geometry is stationary the scalar fields can be time-dependent and such rotating black hole solutions have already been constructed numerically \cite{Herdeiro:2014goa, Collodel:2020gyp}. Even more, based on numerical solutions it was shown that when the scalar fields are time-dependent (periodic in time), there exist fully regular multi-black hole solutions which have stationary, axisymmetric, and asymptotically flat spacetime \cite{Herdeiro:2023roz}. At this stage, it is not clear how to pose and attack the classification problem in its full generality.

As one should expect, black hole classification depends strongly on the particular scalar-tensor theory. For some scalar-tensor theories, a complete or partial classification of the stationary black hole solutions in the vacuum sector is possible, eventually under imposing certain restrictive conditions on the spacetime and/or on the scalar fields. For others, the black hole classification problem is extremely difficult. Examples of such scalar-tensor theories are the theories exhibiting the so-called spontaneous scalarization and in particular the scalar-Gauss-Bonnet gravity \cite{Doneva:2022ewd}. Specific challenges that arise will be demonstrated in explicit examples in the last section of this review.

Our first purpose here is to present most of the known no-hair results, and where possible, new ones, as well as to demonstrate some of the difficulties that appear in our attempts to classify the black holes within scalar-tensor gravity. Our second purpose is to demonstrate some of the mathematical techniques for proving no-hair theorems which have their own independent value and could eventually be useful in the future developments of the subject. In our considerations, we shall restrict to four-dimensional black hole solutions with a connected horizon in asymptotically flat spacetimes belonging to the vacuum sector of scalar-tensor gravity, i.e. without the presence of other fields different from the scalar fields and in particular without the presence of gauge fields. To make the review self-contained we also present the vacuum black hole uniqueness theorems in GR and their proofs. 

The paper is organized as follows. In Section 2 we briefly review the scalar-tensor gravity. Section 3 is devoted to the no-hair (uniqueness) theorems in GR and the mathematical techniques for their proofs. In Section 4 we review the non-hair theorems in classical scalar-tensor theories with single and multiple scalar fields. The necessary mathematical techniques are also discussed and the proofs are presented in detail. In Sections 5 and 6 we focus on the known no-hair theorems and their proof in scalar-tensor theories with a non-canonical scalar field and in Horndeski gravity. The last section 7 discusses some nontrivial challenges in black hole classification in theories exhibiting spontaneous scalarization.

\section{Scalar-tensor theories of gravity} 

In scalar-tensor theories, as we mentioned, the gravitational interaction is mediated not only by the spacetime metric but also by one or more dynamical scalar fields. They constitute the most studied and the most popular theories beyond GR, attracting the efforts of many researchers. 
The roots of the scalar-tensor theories can be traced back to the Kaluza-Klein theory which is in fact the ancestor of the scalar-tensor gravity \cite{Kaluza_1921} - \cite{Duff:1986hr}. Although scalar-tensor theories are theoretically in deep connection with the Kaluza--Klein theory and the higher dimensional gravity as a whole, in certain cases they can also be built on a different base related to solely physical principles \cite{Brans:1961sx, Dicke:1961gz}. 

Scalar-tensor theories come in different varieties. Here we will focus on those scalar-tensor theories that are the most popular and have been the focus of scientists for the last decades.
We will start with the classical scalar-tensor theories and then move on to their more modern versions that have attracted interest in recent years.

\subsection{Classical scalar-tensor theories with a single scalar field}

In the classical scalar-tensor theories, the gravitational sector is augmented with an additional dynamical scalar field $\Phi$, through a non-minimal coupling to the Ricci scalar. From a physical point of view $\Phi^{-1}$ plays the role of a dynamical gravitational ``constant''. For reasons that will become clear a bit later, in writing the scalar-tensor action it is convenient to denote the spacetime metric by ${\tilde g}_{\mu\nu}$ and the quantities associated with it also with a tilde. The most general classical scalar-tensor theories with a single scalar field for which the action is at most quadratic in derivatives of the fields were proposed by Bergmann and Wagoner \cite{Bergmann:1968ve, Wagoner:1970vr} (see also \cite{Will:2018bme}). The explicit form of the Bergmann-Wagoner action is
\begin{equation}
	\label{GBDA}
	S = {1\over 16\pi}\int \! d^4x\sqrt{-{\tilde g}}\left(\!\!\Phi {\tilde R} -
	{\omega(\Phi) \over \Phi}{\tilde g}^{\mu\nu}{\tilde \nabla}_{\mu}\Phi{\tilde \nabla}_{\nu}\Phi -
	U(\Phi) \!\!\right) + S_{\text{matter}}(\Psi,{\tilde g}_{\mu\nu}), 
\end{equation}
with $\omega(\Phi)$ and $U(\Phi)$ being arbitrary functions of the scalar field and $S_m(g_{\mu\nu},\psi)$ is the action of the matter fields collectively denoted by $\psi$.

Varying the action with respect to ${\tilde g}_{\mu\nu}$ and $\Phi$ we get the scalar-tensor field equations
\begin{eqnarray}
	&&{\tilde R}_{\mu\nu} - {1\over 2}{\tilde g}_{\mu\nu}{\tilde R} =
	{8\pi \over \Phi} {\tilde T}_{\mu\nu} + {1\over \Phi} \left({\tilde \nabla}_{\mu}{\tilde \nabla}_{\nu}\Phi -
	{\tilde g}_{\mu\nu}{\tilde \nabla}_{\lambda}{\tilde \nabla}^{\lambda}\Phi\right)
	\nonumber \\
	&& \;+ {\omega(\Phi) \over \Phi^2}
	\left(\!\!{\tilde \nabla}_{\mu}\Phi{\tilde \nabla}_{\nu}\Phi -
	{1\over 2}{\tilde g}_{\mu\nu} {\tilde \nabla}_{\lambda}\Phi{\tilde \nabla}^{\lambda}\Phi \!\!\right) \!- \!
	{U(\Phi)\over 2\Phi}{\tilde g}_ {\mu\nu} , \\
	&& {\tilde \nabla}_{\lambda}{\tilde \nabla}^{\lambda}\Phi =
	{1\over 3 + 2\omega(\Phi)} \left(8\pi {\tilde T} -
	\omega^{{\,}\prime}(\Phi){\tilde \nabla}_{\lambda}\Phi{\tilde \nabla}^{\lambda}\Phi 
	\right. \nonumber \\ && \;\; \;\left. + \Phi U^{\prime}(\Phi) - 2U(\Phi)\right), \nonumber 
\end{eqnarray}
with ${\tilde T}_{\mu\nu}$ being the matter energy-momentum tensor. Using the contracted Bianchi identities and the field equations one can show that energy-momentum tensor satisfies
\begin{equation}
	{\tilde \nabla}_{\mu} {\tilde T}^{\mu}_{\nu} = 0.
\end{equation}

\subsection{Einstein frame presentation of the classical scalar-tensor theories}

The above-given scalar-tensor theory field equations are in the so-called Jordan frame. This is the physical frame of the system, where the gravitational ``constant'' is
a dynamical variable and the elementary particle masses are constants. Due to the non-minimal coupling between the scalar field and Ricci scalar in the Jordan frame, there is an entanglement between the scalar and the tensor degrees of freedom. One can introduce a mathematically related frame where the scalar and the tensor  degrees of freedom are disentangled. This is the so-called Einstein frame where the gravitational constant is a constant, and the scalar field changes the masses of the elementary particles \cite{Dicke:1961gz}. The transformation between the physical Jordan and the Einstein frames is done in the following way. We first rescale the scalar field $\Phi$ by extracting from it a multiplicative constant having dimensions of an inverse gravitational constant $\Phi= G^{-1}_{*}{\tilde \Phi}$. Then the action transforms to 
\begin{eqnarray} \label{GSTA}
	S = {1\over 16\pi G_{*}} \int d^4x \sqrt{-{\tilde g}}
	\left({\tilde \Phi}{\tilde R} -
	{\omega({\tilde \Phi})\over {\tilde \Phi}}
	{\tilde g}^{\mu\nu}{\tilde \nabla}_{\mu}{\tilde \Phi}{\tilde \nabla}_{\nu}{\tilde \Phi}
	- {\tilde U}({\tilde \Phi}) \right) + S_{\text{matter}}(\Psi,{\tilde g}_{\mu\nu}),
\end{eqnarray}
where ${\tilde U}=G_{*} U $. The second step is to conformally transform the metric \begin{eqnarray}
	g_{\mu\nu} = {\tilde \Phi} {\tilde g}_{\mu\nu},
\end{eqnarray}
and introduce a new scalar field 
\begin{eqnarray}
	\varphi = \int^{\Phi} d\zeta
	{\sqrt{3 + 2\omega(\zeta)} \over 2\zeta }.
\end{eqnarray}
Then we reach the following form of the Einstein frame action
\begin{eqnarray}
	\label{AAF}
	S={1\over 16\pi G_{*}} \int d^4x \sqrt{-g}
	\left(R - 2 g^{\mu\nu}
	\partial_{\mu}\varphi\partial_{\nu}\varphi -
	V(\varphi)\right) + 
	S_{\text{matter}}[A^2(\varphi)g_{\mu\nu},\psi].
\end{eqnarray}
where $\nabla_{\mu}$ and $R$ are the covariant derivative and Ricci tensor with respect to the Einstein frame metric $g_{\mu\nu}$ and $A^2 (\varphi) = {\tilde \Phi}^{-1}(\varphi)$. The Einstein frame potential is defined as 
\begin{eqnarray}
	V(\varphi) = A^4(\varphi) U({\tilde \Phi}(\varphi)). 
\end{eqnarray}
The field equations in this frame are obtained by varying the action (\ref{AAF}) with respect to $g_{\mu\nu}$ and $\varphi$:
\begin{eqnarray}
	\label{FEAF}
	&&R_{\mu\nu}- \frac{1}{2}g_{\mu\nu}R =8\pi G_{*} T_{\mu\nu} + 2\nabla_{\mu}\varphi\nabla_{\nu}\varphi - g_{\mu\nu} \nabla_{\sigma}\varphi\nabla^{\sigma}\varphi
	-{1\over 2}g_{\mu\nu}V(\varphi) ,\notag \\
	&&\nabla_{\mu}\nabla^{\mu}\varphi =
	- 4\pi G_{*} \alpha(\varphi) T + {1\over 4}V^{{\,}\prime}(\varphi),
\end{eqnarray}
where $T$ is the trance of the Einstein frame energy-momentum tensor, $T=g^{\mu\nu}T_{\mu\nu}$, given by 
\begin{eqnarray}
	T^{\mu\nu}= - \frac{2}{\sqrt{-g}} \frac{\delta S_{Matter}(A^2(\varphi)g_{\mu\nu},\psi)}{\delta g_{\mu\nu}}.	
\end{eqnarray}
The function $\alpha(\varphi)$, defined by
\begin{eqnarray}
	\alpha(\varphi) = A^{-1}(\varphi){dA(\varphi)\over d\varphi},
\end{eqnarray}
describes the interaction strength between the scalar field $\varphi$ and the matter sources in the Einstein frame.
It is straightforward to check that the energy-momentum tensors in the Einstein frame $T_{\mu\nu}$ and in the Jordan frame ${\tilde T}_{\mu\nu}$, are related through the conformal factor $ A(\varphi)$, namely
\begin{eqnarray} 
	T_{\mu\nu} = A^4(\varphi) {\tilde T}_{\mu\nu}.
\end{eqnarray}

The contracted Bianchi identities combined with the field equations (\ref{FEAF}) lead to the following conservation law
\begin{eqnarray}
	\nabla_{\nu}T_{\mu}^{\nu} =
	\alpha(\varphi)T\partial_{\mu}\varphi .
\end{eqnarray}

\subsection{Classical multiscalar theories of gravity}\label{Multi}

The classical scalar-tensor theories with a single scalar field can naturally be extended to theories with several dynamical scalar fields. Such multiscalar theories arise 
after dimensional reduction of the higher dimensional Kaluza-Klein theories and of the (super)string theory \cite{Witten:1981me, Mallik:1984wc, Duff:1986hr}. It is also worth mentioning, that some specific 
higher derivative gravitational theories are also dynamically equivalent to scalar-tensor theories with multiple scalar fields \cite{Wands:1993uu}. 

Here we consider multiscalar theories containing $N$ dynamical scalar fields $\varphi^a$, having an Einstein frame action given by \cite{Damour:1992we}
\begin{eqnarray}\label{AMSTT}
	S = {1\over 16\pi G_{*}} \int d^4x\sqrt{-g}
	\left(R -
	2\gamma_{ab}(\varphi)
	\nabla_{\mu}\varphi^{a}\nabla^{\mu}\varphi^{b} -
	V(\varphi)\right) + \;
	S_{\text{matter}}[A^2(\varphi)g_{\mu\nu},\Psi],
\end{eqnarray}
where $\gamma_{ab}(\varphi)$ is an arbitrary positive define metric in the ``space'' of fields $\varphi^{a}$ and $a,b,c,... = 1,2 ,3,...N $.
Mathematically, the scalar fields $\varphi^a$ can be thought of as coordinates on an $N$-dimensional Riemannian manifold ${\cal E}_N$ (the so-called {\it target space}) with metric $\gamma_{ab}(\varphi)$. 
An important fact we should note is that the action \eqref{AMSTT} is invariant not only under the group of the spacetime diffeomorphisms but also under the group of the target-space diffeomorphisms. 

The Einstein frame field equations corresponding to the action (\ref{AMSTT}) are 
\begin{eqnarray}
	&&R_{\mu\nu} - \frac{1}{2}g_{\mu\nu}R = 8\pi G_{*} T_{\mu\nu} + 2{\gamma}_{ab}(\varphi) \nabla_{\mu}\varphi^{a}\nabla_{\nu}\varphi^{b}
	\notag \\ 
	&& - g_{\mu\nu} \gamma_{ab}(\varphi)\nabla_{\sigma}\varphi^a\nabla^{\sigma}\varphi^b - {1\over 2}g_{\mu\nu}V(\varphi) ,
	\\ 
	&&\nabla_{\mu}\nabla^{\mu}\varphi^{a}
	+ {\gamma^{\,a}}_{bc}(\varphi)
	\nabla_{\mu}\varphi^{b}\nabla^{\mu}\varphi^{c}
	=
	- 4\pi G_{*}\alpha^{a}(\varphi) T + {1\over 4}V^{a}(\varphi), \notag
\end{eqnarray}
where ${\gamma^{\, a}}_{bc}(\varphi)$ are the target space metric Christoffel symbols and the indices $a,b,c...$ are raised and lowered with the metric ${\gamma}_{ab}(\varphi)$. The coupling function $\alpha(\varphi)$ and the potential derivative generalize, in the case of multiple scalar fields, to
\begin{eqnarray}
	\alpha^{a}(\varphi) =
	{\gamma}^{ab}(\varphi) A^{-1}(\varphi){\partial A(\varphi) \over \partial \varphi^{b}}, \; \;\;\;
	V^{a}(\varphi) =
	{\gamma}^{ab}(\varphi){\partial V(\varphi) \over
		\partial \varphi^{b}} .
\end{eqnarray}

\subsection{Extended scalar-tensor theories of gravity}\label{ESTTG}
In general, we can construct gravitational theories possessing a scalar field (or multiple scalar fields) coupled to the spacetime curvature in a more involved way compared to the classical scalar-tensor theories. 
We shall call them {\it extended scalar-tensor theories}. They are naturally related to higher dimensional gravity \cite{Charmousis:2008kc}. 
For example, higher dimensional theories different from the higher dimensional Einstein gravity, can give rise to non-trivial couplings of the scalar field(s) to the curvature invariants of second order in the curvature (especially the Gauss-Bonnet invariant) under dimensional reduction. It is interesting to point out that the low-energy effective theory of heterotic strings also predicts a modification of Einstein's gravity with a scalar field exponentially coupled to the Gauss-Bonnet
 invariant. This theory is typically referred to as Einstein-dilaton-Gauss-Bonnet gravity \cite{Metsaev:1987zx}.

With this motivation in mind, it is natural to consider extended scalar-tensor theories possessing a scalar field coupled to second-order curvature invariants. There are only four independent such curvature t invariants:
\begin{eqnarray}
	R^2, \;\; R_{\mu\nu}R^{\mu\nu}, \; \; R_{\mu\nu\alpha\beta}R^{\mu\nu\alpha\beta}, \;\; {\cal R}*{\cal R} ,
\end{eqnarray} 
where the last term is the Chern-Pontryagin invariant defined by 
\begin{eqnarray}\label{CSI}
	{ {\cal R}*{\cal R} } = \frac{1}{2}\epsilon^{\mu\nu\alpha\beta} R_{\mu\nu\rho\sigma} R_{\alpha\beta}^{\;\;\;\;\;\rho\sigma}.
\end{eqnarray}	

Therefore, let us consider the following Einstein frame action of an extended scalar-tensor theory, having a single scalar field \cite{Yunes:2011we, Pani:2011gy}
\begin{eqnarray}\label{AESTT}
	S &=& \frac{1}{16\pi G}\int d^4x\sqrt{-g} \left[R - 2\nabla_{\mu}\varphi\nabla^{\mu}\varphi - V(\varphi) + f_1(\varphi)R^2 + \; f_2(\varphi) R_{\mu\nu}R^{\mu\nu}
		\nonumber \right. \\ && \left. + f_{3}(\varphi) R_{\mu\nu\alpha\beta}R^{\mu\nu\alpha\beta} + f_{4}(\varphi) {\cal R}*{\cal R} \right] 
	 + \; S_{matter}(A^2(\varphi) g_{\mu\nu},\Psi),
\end{eqnarray}
where the coupling functions between the scalar field and the curvature invariants are denoted by $f_i(\varphi)$. All of them have dimensions of a length
squared, i.e. of an inverse curvature. Therefore, the extended scalar-tensor theories require the introduction of new fundamental length scales. 

Among the most popular representatives are the dynamical Chern-Simons gravity and the scalar-Gauss-Bonnet gravity. The former one is constructed by setting $f_1=f_2=f_3=0$ and keeping only the coupling to the Chern-Pontryagin invariant. Namely, we set $f_4(\varphi)= \alpha_{CS}\varphi$ with $\alpha_{CS}$ being a coupling parameter with dimension of length squared. Additionally, the formulations of dynamical Chern-Simons gravity typically do not consider matter coupling, namely $A^2(\varphi)=1$. Thus, the Chern-Simons theory action takes the form \cite{Alexander:2009tp}
\begin{eqnarray}
	S_{CS} = \frac{1}{16\pi G}\int d^4x\sqrt{-g} \left[R - 2\nabla_{\mu}\varphi\nabla^{\mu}\varphi - V(\varphi) + \alpha_{SC} \varphi {\cal R}*{\cal R} \right] 
 + S_{matter}(g_{\mu\nu},\Psi).
\end{eqnarray}
It should be stressed that in the dynamical Chern-Simons gravity, $\varphi$ is in fact a pseudo-scalar (axion-like) field. The theory is parity-violating -- the deviations from general relativity
occur only for systems that violate parity through the presence of a preferred axis. Examples of such systems are isolated rotating black holes and neutron stars. 

The second very popular example is the scalar-Gauss-Bonnet gravity. It is constructed by setting the Chern-Pontryagin invariant coupling to zero ($f_4=0$) and taking a special combination of the rest of the curvature invariants corresponding to the Gauss-Bonnet scalar, namely $f_2(\varphi)=-4f_1(\varphi)$ and $f_{3}(\varphi)=f_1(\varphi)$. Assuming a standard form of $f_1(\varphi)=\lambda^2 f(\varphi)$, where $\lambda$ is a coupling constant (parameter) with a dimension of length, the general action (\ref{AESTT}) reduced to the scalar-Gauss-Bonnet gravity one:
\begin{eqnarray}\label{ASGBG}
	S_{GB} &=& \frac{1}{16\pi G}\int d^4x\sqrt{-g} \left[R - 2\nabla_{\mu}\varphi\nabla^{\mu}\varphi - V(\varphi) + \lambda^2 f(\varphi) {\cal R}^2_{\,\, GB} \right] 
	\nonumber \\ && + \; \; S_{matter}(A^2(\varphi)g_{\mu\nu},\Psi).
\end{eqnarray}
Here ${\cal R}^2_{\,\, GB}=R^2 - R_{\mu\nu}R^{\mu\nu} + R_{\mu\nu\alpha\beta}R^{\mu\nu\alpha\beta}$ denotes the Gauss-Bonnet invariant. Most of the subclass of scalar-Gauss-Bonnet theories, intensively studied in the literature, assume no coupling to matter $A(\varphi)=1$.

The action (\ref{AESTT}), in its general form, yields higher-order field equations that are prone to the Ostrogradski instability and the appearance of ghosts. Fortunately, there is one exception, namely scalar-Gauss-Bonnet gravity. In this special combination of $f_i$ couplings, the field equations are of second order as in general relativity and the theory is free from ghosts.

By varying the action (\ref{ASGBG}) with respect to the spacetime metric $g_{\mu\nu}$ and the scalar 
field $\varphi$, one can obtain the scalar-Gauss-Bonnet vacuum field equations:
\begin{eqnarray}\label{FESGBG_1}
	&&	R_{\mu\nu} - \frac{1}{2}R g_{\mu\nu} + \Gamma_{\mu\nu}= 8\pi G_* T_{\mu\nu} + 2\nabla_\mu\varphi\nabla_\nu\varphi - g_{\mu\nu} \nabla_\alpha\varphi \nabla^\alpha\varphi 
	- \frac{1}{2} g_{\mu\nu}V(\varphi), \\
	&&	\nabla_\alpha\nabla^\alpha\varphi = - 4\pi G_* \alpha(\varphi)T + \frac{1}{4} \frac{dV(\varphi)}{d\varphi} - \frac{\lambda^2}{4} \frac{df(\varphi)}{d\varphi} {\cal R}^2_{\;\;GB}, \notag
\end{eqnarray}
where the symmetric tensor $\Gamma_{\mu\nu}$ is defined by 
\begin{eqnarray}\label{FESGBG_2}
	\Gamma_{\mu\nu}&=& - R(\nabla_\mu\Psi_{\nu} + \nabla_\nu\Psi_{\mu} ) - 4\nabla^\alpha\Psi_{\alpha}\left(R_{\mu\nu} - \frac{1}{2}R g_{\mu\nu}\right) + \; 	4R_{\mu\alpha}\nabla^\alpha\Psi_{\nu} + 4R_{\nu\alpha}\nabla^\alpha\Psi_{\mu} 
\nonumber \\
&& 	- \; 4 g_{\mu\nu} R^{\alpha\beta}\nabla_\alpha\Psi_{\beta} + \; 4 R^{\beta}_{\;\mu\alpha\nu}\nabla^\alpha\Psi_{\beta} 
\end{eqnarray} 
with 
\begin{eqnarray}\label{FESGBG_3}
	\Psi_{\mu}= \lambda^2 \frac{df(\varphi)}{d\varphi}\nabla_\mu\varphi .
\end{eqnarray}

The action \eqref{ASGBG} is invariant under the following transformation $f(\varphi)\to f(\varphi) + constant$ which is a consequence of the fact that ${\cal R}^2_{\,\, GB}$ is a topological invariant. The same transformation naturally leaves also the field equations \eqref{FESGBG_1} and \eqref{FESGBG_2} invariant since only derivatives of the coupling function $f(\varphi)$ enter there.

\subsection{Horndeski theories}\label{Horndeski}

The most general scalar-tensor theory with a single scalar field and second-order field equations is the so-called Horndeski gravity \cite{Horndeski:1974wa, Deffayet:2009mn}. The vacuum action given by
\begin{eqnarray} \label{Hordeski_Action}
	S_H = \frac{1}{16\pi G_*} \int d^4x\sqrt{-g}\left[ {\cal L}_2+ {\cal L}_3+ {\cal L}_4 + {\cal L}_5\right]
\end{eqnarray}
with 
\begin{eqnarray}\label{HorndeskiFUN}
	&&{\cal L}_2= F(\varphi, K), \notag\\
	&&{\cal L}_3 = - G_{3}(\varphi,K)\nabla_{\mu}\nabla^{\mu}\varphi , \notag\\
	&&{\cal L}_4 =G_{4}(\varphi, K) R + \partial_{K}G_4(\varphi,K)[(\nabla_{\mu}\nabla^{\mu}\varphi)^2 - \nabla_{\mu}\nabla_{\nu}\varphi \nabla^{\mu}\nabla^{\nu}\varphi], \notag \\
	&& {\cal L}_5 = G_5(\varphi,K)G^{\mu\nu}\nabla_{\mu}\nabla_{\nu}\varphi \notag \\ 
	&&- \frac{2}{3} \partial_{K}G_{5}(\varphi,K) [(\Box\varphi)^3 -3\Box\varphi \nabla_{\mu}\nabla_{\nu}\varphi \nabla^{\mu}\nabla^{\nu}\varphi 
	+ 2\nabla_{\mu}\nabla_{\nu}\varphi \nabla^{\mu}\nabla_{\alpha}\varphi \nabla^{\nu}\nabla^{\alpha}\varphi], \notag
\end{eqnarray} 
where $F$ and $G_i$ are functions of the scalar field $\varphi$ and its kinetic energy $K=-2\nabla_{\mu}\varphi \nabla^{\mu}\varphi$, and $G_{\mu\nu}$ is the Einstein tensor. 

It is not difficult to see the classical scalar-tensor theories are a subclass of the Horndski gravity. The scalar-Gauss-Bonnet gravity itself is also contained in the Hordenski action for some rather specific choice of the functions $F$ and $G_i$ (see \cite{Kobayashi:2011nu}). Other particular cases of the Horndeski
theory which attracted interest in the literature are briefly discussed below.

A subclass of Horndeski theories is the scalar-tensor gravity with a non-canonical scalar field having an action of the form 
\begin{eqnarray}\label{HordeskiNCSF}
	S = \frac{1}{16\pi G_*} \int d^4x\sqrt{-g}\left[ R + F(\varphi,K) \right].
\end{eqnarray}
This theory includes the K-essence models for $F(\varphi,K)=f(\varphi)F(K)$. The canonical scalar field case is recovered for $F(\varphi,K)=K - V(\varphi)$. The main motivation behind
these models is related to cosmology -- they were first introduced in \cite{Armendariz-Picon:1999hyi} as a new model of inflation and
later studied as a dark energy model \cite{Chiba:1999ka, Armendariz-Picon:2000nqq}. The power of K-essence models is kept in the fact that with an appropriately chosen $F(\varphi,X)$ one can construct an accelerated expansion cosmology. The acceleration will be driven by the kinetic rather than the scalar field potential energy. These theories are an alternative to the quintessence models where the acceleration occurs due to the scalar field slowrolling down its potential.

Another interesting class is the scalar-tensor gravity with a kinetic coupling to the Einstein tensor given in vacuum by the action 
\begin{eqnarray}
	S = \frac{1}{16\pi G_*} \int d^4x\sqrt{-g}\left[ R + K - V(\varphi) + a G^{\mu\nu}\nabla_{\mu}\varphi \nabla_{\nu}\varphi \right] 
\end{eqnarray}
where $a$ is a parameter with dimension of length squared.

\section{Black holes in General Relativity}\label{BHGR}

Here we present some of the basic and most important results concerning the stationary and asymptotically flat black holes in vacuum general relativity.
We also present the necessary mathematical techniques and constructions which have their independent value as powerful mathematical tools in the gravitational theories and which will be used in 
the next sections in proving no-hair theorems in the scalar-tensor theories. 

We shall assume that the matter energy-momentum satisfies some natural energy conditions such as the weak energy condition, i.e. $T_{\mu\nu}u^{\mu} u^{\nu}\ge 0$ for every future-pointing timelike vector $u^{\mu}$. As it is well-known, this condition means that the energy density measured by any observer on a timelike curve has to be non-negative. A limiting case of the weak energy condition is the null energy condition according to which $T_{\mu\nu}k^{\mu} k^{\nu}\ge 0$ for every future-pointing null vector $k^{\mu}$. Another energy condition needed in the theory of stationary black holes is the average null energy condition which requires $\int d\lambda T_{\mu\nu}k^{\mu} k^{\nu} \ge 0 $ along every inextendible null geodesic with affine parameter $\lambda$ and corresponding tangent vector $k^{\mu}$. The null energy condition obviously implies the average null energy condition.

We denote by $\xi$ the stationary Killing vector, ${\cal L}_{\xi} g_{\mu\nu} = 0$, which is asymptotically timelike and is normalized so that $g_{\mu\nu}\xi^{\mu}{\xi^{\nu}}=-1$ at infinity. The horizon of the black hole will be denoted by $H$ and we will assume that $H$ is non-degenerate. We also assume that the domain of outer communication $\left<{\cal M}\right>$ is globally hyperbolic. The topological censorship theorem \cite{Galloway:1999bp}, when the matter satisfies the averaged null energy condition, states that $\left<{\cal M}\right>$ is a simply connected manifold with boundary $\partial\left<{\cal M}\right>=H$. As a consequence of the topological censorship theorem, it also follows that the horizon cross-section ${\cal H}$, i.e. the cross-section of the horizon with a partial Cauchy surface, has ${\mathbb S}^2$ topology. All these results imply that the topology of the domain of outer communications
must be $\left<{\cal M}\right>={\mathbb R}\times ({\mathbb R}^3 \setminus B)$ where $({\mathbb R}^3 \setminus B)$ is the complement of a 3-dimensional open ball $B$ in Euclidean space ${\mathbb R}^3$. 

A key result in the black hole theory within General Relativity is the so-called  ``rigidity theorem'' \cite{Hawking:1973uf, Chrusciel:1996bj, Friedrich:1998wq}, which relates the global concept of
the event horizon to the local notion of the Killing horizon. Let us recall the definition of a Killing horizon associated with the Killing vector ${\cal K}$. A null hyper-surface that coincides with the connected component of the set 
\begin{eqnarray}
	N[{\cal K}]=\{g_{\mu\nu}{\cal K}^{\mu} {\cal K}^{\nu}=0 ,\;\; {\cal K}\ne 0\}
\end{eqnarray}
is called a Killing horizon associated with the Killing vector ${\cal K}$. 

Having defined a Killing horizon we can present the rigidity theorem:

\medskip
\noindent

{\bf Theorem } {\it Let $({\cal M},g)$ be an analytic, asymptotically flat, stationary black hole space-time. Furthermore, let the energy-momentum tensor of matter satisfy the weak energy condition. Then, the event horizon is a Killing horizon, i.e. there exists a Killing field ${\cal K}$ which has an associated Killing horizon coinciding with the event horizon. The Killing field ${\cal K}$ either coincides with the stationary Killing field $\xi$ and the spacetime is static, or space-time admits at least one axial Killing field $\eta$ (i.e. with periodic orbits) such that ${\cal K}=\xi + \Omega_{H}\eta$, where $\Omega_{H}=const$. }

\medskip
\noindent

 The rigidity theorem states that any stationary black hole is either static or axially symmetric and rotating with a constant angular velocity $\Omega_{H}$. The Killing field ${\cal K}$ is tangent and normal to the null generators of the horizon and 
\begin{eqnarray}
	\left. g_{\mu\nu}{\cal K}^{\mu}\eta^{\nu}\right|_{H}=0.
\end{eqnarray}

The surface gravity $\kappa$ of the black hole horizon is given by
\begin{eqnarray}
	\kappa = \lim_{H}\frac{\nabla_{\mu}n\nabla^{\mu}n}{n}
\end{eqnarray}
where $n=\nabla^{\mu}{\cal K}^{\nu}\nabla_{\mu}{\cal K}_{\nu}$. Let us also recall the well-known fact that the surface gravity $\kappa$ is a constant on the (Killing) horizon. The horizon non-degeneracy condition means $\kappa>0$.

\subsection{Static vacuum black holes}\label{SBHGR}

The Schwarzschild solution to the vacuum Einstein equations is the simplest black hole solution which is asymptotically flat, spherically symmetric, and static in the domain of outer communication. In the standard coordinates it is given by 
\begin{equation}\label{SBHS}
	ds^2= g_{\mu\nu}dx^\mu dx^\nu = - (1- \frac{2M}{r})dt^2 + \frac{dr^2}{ 1- \frac{2M}{r}} + r^2 d\Omega^2
\end{equation}	
where $d\Omega^2=d\theta^2 + \sin^2\theta d\phi^2$ is the metric on the unit sphere and $M$ is the mass of the black hole. The event horizon lays 
at $r=2M$. The following key uniqueness theorem holds
\medskip
\noindent

{\bf Theorem} {\it The Schwarzschild metric represents the only vacuum black hole solution with a regular event horizon and static, 
	asymptotically flat domain of outer communications.}

\medskip
\noindent

We shall sketch the proof of this important result following the original paper \cite{Bunting_1987}. The spacetime metric in the domain of outer communication assumes the standard form
\begin{eqnarray}\label{SMETRIC}
	ds^2= -N^2 dt^2 + g^{(3)}_{ij}dx^i dx^j ,
\end{eqnarray} 
where $g^{(3)}_{ij}$ the Riemannian metric on the 3-dimensional hypersurface $\Sigma$ orthogonal to $\xi=\frac{\partial}{\partial t}$. The 3-dimensional manifold $\Sigma$ is a manifold with a boundary $\partial\Sigma = {\cal H}$. The asymptotic flatness means that outside a compact set, $\Sigma$ is diffeomorphic to ${\mathbb R}^3/{\bar B}$ where ${\bar B}$ is a closed ball centered at the origin of ${\mathbb R}^3$. In addition $N$ and the 3-dimensional metric $g^{(3)}_{ij}$ have the following asymptotics
\begin{eqnarray}
	N=1 - \frac{M}{r} + O(\frac{1}{r^2}), \;\;\;\; g^{(3)}_{ij}= (1 + \frac{M}{r})\delta_{ij} + O(\frac{1}{r^2})
\end{eqnarray} 
in the standard coordinates $x^i$ in ${\mathbb R}^3$ where $r=\sqrt{\delta_{ij}x^ix^j}$ and $M$ is the ADM mass. In the mathematical literature $(\Sigma, g^{(3)}_{ij})$ is called {\it asymptotically flat 3-dimensional Riemannian manifold with mass $M$}. 

The dimensionally reduced Einstein equations on $\Sigma$
take the form 
\begin{eqnarray}
	&&R^{(3)}_{ij} = N^{-1} D_{i}D_{j}N, \\
	&&D^{i}D_{i}N =0,
\end{eqnarray} 
where $D_i$ is the covariant derivative of the 3-dimensional manifold $(\Sigma, g^{(3)}_{ij})$ and $R^{(3)}_{ij}$ is its Ricci tensor. Hence it is easy to see that the scalar curvature $R^{(3)}$ of $(\Sigma, g^{(3)}_{ij})$ is vanishing.

The strategy of the proof is to use the positive mass theorem \cite{Schon:1979rg, Witten:1981mf} and more precisely its Riemannian version, namely

\medskip
\noindent

{\bf Theorem} {\it Let $({\hat \Sigma}, {\hat g}^{(3)}_{ij})$ be an asymptotically flat complete, orientable 3-dimensional Riemannian
	manifold with non-negative scalar curvature ${\hat R}\ge 0$ and vanishing mass, ${\hat M}=0$ . Then $({\hat \Sigma}, {\hat g}^{(3)}_{ij})$ is isometric to $({\mathbb R}^{3},\delta_{ij}$)}.

\medskip
\noindent

In order to reduce the problem to the positive mass theorem we need to consider conformal transformations of $(\Sigma, g^{(3)})$, namely 
\begin{eqnarray}
	{\hat g}^{(3)}_{ij}=\Omega^2 g^{(3)}_{ij}
\end{eqnarray} 
with appropriate conformal factors. Let us recall that under such conformal transformations, the Ricci scalar transforms according to the law
\begin{eqnarray}\label{CT}
	\Omega^4{\hat R}^{(3)} = \Omega^2 R^{(3)} - 4\Omega D_{i}D^{i}\Omega + 2D_{i}\Omega D^{i}\Omega.
\end{eqnarray} 

In order to reduce the problem to the positive mass theorem, we consider two copies of $\Sigma$, ${\hat \Sigma}_{+}$ and ${\hat \Sigma}_{-}$, with boundaries ${\cal H}_{+}$ and ${\cal H}_{-}$, respectively, and metrics 
\begin{eqnarray}\label{CFGR}
	g^{(3,\pm)}_{ij} = \Omega^2_{\pm} g^{(3)}_{ij}, \;\;\; \Omega_{\pm}= \frac{1}{4}(1\pm N)^2 .	
\end{eqnarray}
The next step is to show that the above conformal transformations are regular. The conformal factor $\Omega_{+}$ is obviously regular and $\Omega_{+}\ge \frac{1}{4}$ since the domain of outer communications is static and $N>0$ on $\Sigma$. The case of $\Omega_{-}$ is more subtle. We notice that $(1-N)$ is a harmonic function since $N$ is harmonic ($D^{i}D_{i}N =0$). In addition, asymptotically we have $(1-N)=\frac{M}{r} + O(1/r^2)$. So in the asymptotic region $(1-N)$ is positive. Now on the boundary $\partial\Sigma={\cal H}$, relative to the outward normal $n$ of ${\cal H}$, we have $\frac{\partial (1-N)}{\partial n}=-\frac{\partial N}{\partial n}=-\kappa$ with $\kappa$ being the surface gravity. Since the horizon is regular, i.e. $\kappa>0$, we find $\frac{\partial (1-N)}{\partial n}<0$.
Then according to the maximum principle for elliptic partial differential equations, the minimum of $(1-N)$ can not be reached on $\Sigma$. In other words $\Omega_{-}\ne 0 $ on $\Sigma$.

We proceed further by considering the manifold ${\hat \Sigma}_{+\cup -}= \Sigma_{+}\cup \Sigma_{-}$ 
obtained by pasting two copies of $\Sigma$ along the common boundary ${\cal H}$. ${\hat \Sigma}_{+\cup -}$ is supplemented with a metric ${\hat g^{(3)}}_{ij}$
such that ${\hat g^{(3)}}_{ij}|_{\Sigma_{\pm}}=g^{(3\pm)}_{ij}$. Apparently the metric is continuous across ${\cal H}$ since $N=0$ on ${\cal H}$. A direct calculation also shows the second fundamental form also matches continuously on ${\cal H}$. 

By construction, ${\hat \Sigma}_{+\cup -}$ has two asymptotically flat regions related to $\Sigma_{+}$ and $\Sigma_{-}$, respectively.
The asymptotic behaviour of the metric ${\hat g^{(3)}}_{ij}$ in the asymptotic region related to ${\hat \Sigma}_{+}$ can be easily derived and we get 
\begin{eqnarray}
	{\hat g^{(3)}}_{ij} = \delta_{ij} + O(\frac{1}{r^2}). 	
\end{eqnarray}
In other words, the mass ${\hat M}$ of ${\hat \Sigma}_{+\cup -}$ associated with the asymptotic region of $\Sigma_{+}$ is just ${\hat M}=0$. The asymptotic behaviour of the metric ${\hat g^{(3)}}_{ij}$ in the asymptotic region corresponding to $\Sigma_{-}$ can also be found without difficulties and we have
\begin{eqnarray}
	{\hat g^{(3)}}_{ij} = \frac{M^{4}}{16r^4}\delta_{ij} + O(\frac{1}{r^6}). 	
\end{eqnarray}
Using the new coordinates ${\tilde x}^i= \frac{x^i}{r^2}$ we find that, in these new coordinates, the behaviour of the metric is 
\begin{eqnarray}
	{\hat g^{(3)}}(\frac{\partial}{\partial {\tilde x}^i}, \frac{\partial}{\partial {\tilde x}^j}) = \frac{M^{4}}{16}\delta_{ij} + O({\tilde r}^2) 	
\end{eqnarray}
as ${\tilde r}\to 0$ where ${\tilde r}^2=\delta_{ij}{\tilde x}^i{\tilde x}^j$. This shows that we can add the infinite point $\infty$ at ${\tilde r}=0$ so that the constructed in this way manifold ${\hat \Sigma}={\hat \Sigma}_{+\cup -} \cup \infty = \Sigma_{+}\cup \Sigma_{-} \cup \infty$ is (sufficiently) regular. By
construction ${\hat \Sigma}$ is geodesically complete and with only one asymptotically flat region and a vanishing mass, ${\hat M}=0$. 

What remains to be done is to calculate the scalar curvature of $({\hat \Sigma},{\hat g^{(3)}}_{ij})$. This can be done by substituting $\Omega_{\pm}=\frac{1}{4}(1\pm N)^2$ into the well-known formula for the transformation of the Ricci scalar under conformal transformations of the metric. Doing so we find 
\begin{eqnarray}
	\Omega^4{\hat R}^{(3)} = \Omega^2_{\pm} R^{(3)} \mp 2\Omega_{\pm}(1 \pm N)\Omega D_{i}D^{i}N =0,
\end{eqnarray} 
where we have taken into account that $ R^{(3)}=0$ and $D_{i}D^{i}N =0$. 

Summarizing, we have shown that $({\hat \Sigma},{\hat g^{(3)}}_{ij})$ is asymptotically flat, complete, orientable 3-dimensional Riemannian
manifold with zero scalar curvature and vanishing mass. Then, as follows from the Riemannian version of the positive mass theorem, we conclude that 
$({\hat \Sigma},{\hat g^{(3)}}_{ij})$ is isometric to $({\mathbb R}^{3},\delta_{ij}$). This in turn shows that the metric ${\hat g^{(3)}}_{ij}$ is conformally flat and 
obviously the same applies to $g^{(3)}_{ij}$, namely 
\begin{eqnarray}
	g^{(3)}_{ij}= \left(\frac{1+ N}{2}\right)^{-4}\delta_{ij}. 
\end{eqnarray} 

The theorem now simply follows by straightforward integration of the filed equations in a spherical coordinate system. Doing so we find the Schwarzschild solution presented in isotropic spherical coordinates:
\begin{eqnarray}
	N= \frac{1 -\frac{M}{2r}}{1 +\frac{M}{2r}}
\end{eqnarray} 
and 
\begin{eqnarray}
	ds^2= -\left(\frac{1 -\frac{M}{2r}}{1 +\frac{M}{2r}}\right)^2 + \left(1 +\frac{M}{2r}\right)^2\left(dr^2 + r^2d\theta^2 + r^2\sin^2\theta d\phi^2 \right).
\end{eqnarray} 
The transition to the standard Schwarzschild coordinates is given by $(1+ \frac{M}{2r})^2 r \to r $. This completes the proof.

\subsection{Rotating vacuum black holes}\label{RBHGR}

The rotating vacuum black holes in general relativity are described by the two-parameter Kerr solution. The Kerr black hole solution has a stationary, axisymmetric, and asymptotically flat domain of outer communications. The explicit form of the Kerr metric written in the standard Boyer-Lindquist coordinates is \cite{Misner:1973prb, Wald:1984}
\begin{eqnarray}\label{KBHS}
	ds^2= - \frac{\Delta -a^2 \sin^2\theta}{\Sigma} dt^2 - 2a\sin^2\theta \frac{r^2 + a^2 - \Delta}{\Sigma} dt d\phi \nonumber \\+
	\left[\frac{(r^2 + a^2)^2 - \Delta a^2 \sin^2\theta}{\Sigma}\right]\sin^2\theta d\phi^2 + \frac{\Sigma}{\Delta} dr^2 + \Sigma d\theta^2
\end{eqnarray}	
with $\Delta=r^2 + a^2 - 2Mr$ and $\Sigma=r^2 + a^2\cos^2\theta$ for $M^2\ge a^2$. This metric describes a rotating black hole with mass $M$ and angular momentum $J=Ma$. The bigger root of $\Delta=0$ corresponds to the event horizon $r_{+}=M +\sqrt{M^2 -a^2} $ while the smaller root $r_{-}=M -\sqrt{M^2 -a^2} $ corresponds to the Cauchy horizon. In addition to the event horizon the Kerr black holes possess also an ergoregion with a boundary given by $g(\frac{\partial}{\partial t},\frac{\partial}{\partial t})=0$ or in explicit form $r_{ergo}=M + \sqrt{M^2 - a^2\cos^2\theta}$. The angular velocity of the black hole horizon is 
\begin{equation}
	\Omega_H = \frac{a}{r^2_{+} +a^2},
\end{equation}
while the surface gravity $\kappa$ and the horizon area ${\cal A}_{H}$ are given by 
\begin{eqnarray}\label{KBHSG}
	\kappa = \frac{1}{2M} \frac{\sqrt{M^4 -J^2}}{M^2 + \sqrt{M^4 -J^2}}, \;\;\; {\cal A}_{H}=8\pi \left(M^2 + \sqrt{M^4 - J^2} \right).
\end{eqnarray}	
Obviously, the metric of Kerr reduces to the Schwarzschild one for $a=0$. 

The Kerr solution is the only solution describing rotating black holes in the vacuum GR. More precisely, we have the following 

\medskip
\noindent

{\bf Theorem} {\it The Kerr metric with parameters $M$ and $a=J/M$ is the only black hole solution with $M^2\ge a^2$, regular event horizon and stationary, asymptotically flat domain of outer communications.} 

\medskip
\noindent

The proof of this fundamental theorem was given in the classical papers \cite{Carter:1971zc, Robinson:1975bv, Mazur:1982db, Bunting_1983} and can also be found in the book \cite{Heusler_1996}. We shall sketch a 
 proof of the above uniqueness theorem combining the classical approach with the approach used in higher dimensions \cite{Hollands:2007aj, Hollands:2008fm}. We will also restrict ourselves to the regular case with $\kappa>0$ since the extreme case with $\kappa=0$ ($M^2=a^2$) requires a bit more technicalities without giving anything more essential.

We focus on the case when the stationary Killing vector $\xi$ is not tangent to the null generators of the 
horizon. Then, as follows from the rigidity theorem, there exists an additional Killing field $\eta$ with a periodic flow of
period $2\pi$, commutes with $\xi$ and has non-empty axis. In other words, the spacetime is also axisymmetric with an isometry group ${\cal G}={\mathbb{R}}\times SO(2)$
with the factor ${\mathbb R}$ corresponding to the stationary symmetry while $SO(2)$ is related to the rotational symmetry.

The vacuum stationary and axisymmetric spacetimes are circular and locally, away from the axis of symmetry and the horizon, they can 
be described as $\left<{\cal M}\right>={{\mathbb R} \times SO(2)}\times {\hat {\cal M}}$ where ${\hat {\cal M}}$ is a 2-dimensional Riemannian manifold orthogonal to ${\mathbb{R}}\times SO(2)$. There is a naturally induced Riemannian metric ${\hat g}_{ij}$ on ${\hat {\cal M}}$, defined as the restriction of the spacetime metric to the vectors orthogonal to the Killing vectors $\xi$ and $\eta$. Then, in adapted coordinates $x^{I}=(t,\phi)$ in which ${\xi}=\frac{\partial}{\partial t}$, ${\eta}=\frac{\partial}{\partial \phi}$, the spacetime metric has the form 
\begin{equation}\label{MCST}
	ds^2 = \Gamma_{IJ}(x^k) dx^Idx^J + {\hat g}_{ij}(x^k)dx^idx^j 
\end{equation}
where $x^k$ are coordinates on ${\hat {\cal M}}$ and $\Gamma_{IJ}$ is the Gram matrix of the Killing fields $K_I$, i.e. 
\begin{equation}
	\Gamma_{IJ}=g_{\mu\nu}K^{\mu}_I K^{\nu}_{J} 
\end{equation}
where $K_1=\xi$ and $K_{2}=\eta$. In explicit form we can write 
\begin{eqnarray}
	\Gamma=\begin{pmatrix} -V & W \\ W & X \end{pmatrix},
\end{eqnarray}
with $V=-g_{tt}$, $W=g_{t\phi}$ and $X=g_{\phi\phi}$.

Since $\det g_{\mu\nu}=\det \Gamma_{IJ} \det g_{ij}$ the determinant of the Gram matrix must be non-positive and we can define
\begin{equation}\label{DEF_rho}
	\rho^2 = -\det \Gamma= VX + W^2. 
\end{equation} 
On the axis of symmetry, i.e. on the points where $\eta$ vanishes, we obviously have $\rho=0$. On the black hole horizon we also have $\rho=0$ since
$g_{\mu\nu}{\cal K}^{\mu}{\cal K}^{\nu}=0$ and $g_{\mu\nu}{\cal K}^{\mu}\eta^{\nu}=0$ there.

From more formal mathematical point of view, ${\hat {\cal M}}$ is in fact the factor space ${\hat {\cal M}}=\left<{\cal M}\right>/{\cal G}$. 
In order to gain some insight into the global structure of the factor space ${\hat {\cal M}}$ we take into account that $\left<{\cal M}\right>$ is topologically ${\mathbb R}\times ({\mathbb R}^3 \setminus B)$ and as a model example we can consider the purely flat spacetime ${\mathbb R}\times ({\mathbb R}^3 \setminus B)$. Within this simplified model, the factorization by the factor ${\mathbb R}$ corresponding to the stationary symmetry can be performed trivially and we find ${\hat {\cal M}}=\left<{\cal M}\right>/{\cal G}= ({\mathbb R}^3 \setminus B)/SO(2)$. It is intuitively clear and this factor is just the a half-plane with boundary  which is a union of the finite segment $\partial B/SO(2)$ (a semicircle), and the semi-infinite northern and southern part of the rotational axis. 

In our case, leaving aside some mathematical subtleties, the things are pretty much the same, and the factor space ${\hat {\cal M}}=\left<{\cal M}\right>/{\cal G}$ is a simply connected 2-dimensional manifold with boundaries and corners homeomorphic to a half-plane. In a more detailed description, the boundary $\partial {\hat {\cal M}}$ consists of the finite interval $I_H$ corresponding to the quotient of the horizon, $I_{H}=H/{\cal G}$ and the two semi-infinite intervals $I_{+}$ and $I_{-}$ corresponding to the axis of $\eta$. The corners correspond to the points where the axis intersects the horizon. The formal and rigorous proof the reader can see \cite{Hollands:2007aj, Hollands:2008fm}.

With the help of the Gram matrix $\Gamma_{IJ}$, one can write the Einstein equations as a system of partial differential equations on
$({\hat {\cal M}}, {\hat g}_{ij})$, namely 
\begin{eqnarray}\label{DRVSASEE}
	{\hat D}^{i}\left(\rho \Gamma^{IL}D_{i}\Gamma_{LJ}\right)=0
\end{eqnarray}	
together with
\begin{eqnarray}
	{\hat R}_{ij} ={\hat D}_{i}{\hat D}_{j}\rho - \frac{1}{4} {\hat D}_{i}\Gamma^{IJ} {\hat D}_{j}\Gamma_{IJ}. 
\end{eqnarray}
Here $\Gamma^{IJ}$ is the inverse of $\Gamma_{IJ}$ and ${\hat R}_{ij}$ is the Ricci tensor associated with $({\hat {\cal M}}, {\hat g}_{ij})$. These equations are well-defined at the points 
in the interior of $\hat {\cal M}$ where $\rho\ne 0$ and the matrix $\Gamma_{IJ}$ is not singular. 

Taking into account that $\Gamma^{IJ}d\Gamma_{IJ}=d\ln\det \Gamma_{IJ}=2d\ln(\rho)$, the trace of equation (\ref{DRVSASEE}) gives 
\begin{eqnarray}
	{\hat D}^{i}{\hat D}_{i}\rho =0.
\end{eqnarray}	
In other words, $\rho$ is a harmonic function on the interior of $\hat {\cal M}$. In addition, $\rho$ vanishes on the boundary of ${\hat {\cal M}}$ and, in the region of ${\hat {\cal M}}$ corresponding to a
neighborhood of infinity of $\left<{\cal M}\right>$ and away from the axis, we have $\rho\to \infty$. Hence, according to the maximum principle for harmonic functions, the function $\rho$ must satisfy $\rho>0$ and ${\hat D}_i\rho \ne 0$ on the interior of ${\hat {\cal M}}$. Thus, $\rho$ can be chosen  as one of the coordinates on $\hat {\cal M}$. A conjugate harmonic scalar field $z$ may then be defined on ${\hat {\cal M}}$ by the equation ${\hat D}_{i}z= \epsilon^{j}_{\;i}{\hat D}_j\rho $, where $\epsilon^{j}_{\;i}$ is the Levi-Civita symbol associated with $({\hat {\cal M}}, {\hat g}_{ij})$. Combining this with the fact ${\hat {\cal M}}$ is homeomorphic to a half-plane, we can conclude that the functions $(\rho,z)$ define global coordinates on ${\hat {\cal M}}$, thus identifying this space with the complex upper half-plane 
\begin{eqnarray}
	{\hat {\cal M}}=\{\zeta= z + i\rho \in {\mathbb C}, \rho\ge 0\}.
\end{eqnarray}	
Using that any 2-dimensional Riemannian manifold is conformally flat, in these
coordinates, the metric ${\hat g}_{ij}$ globally takes the form
\begin{equation}
	{\hat g}_{ij}dx^i dx^j= e^{2\nu(\rho,z)}(d\rho^2 + dz^2).
\end{equation} 
In this way, the spacetime metric may be parametrized as follows 
\begin{eqnarray}
	ds^2= -V(\rho,z)dt^2 + 2W(\rho,z)dtd\phi + X(\rho,z)d\phi^2 + e^{2\nu(\rho,z)}(d\rho^2 + dz^2) .
\end{eqnarray}

Eliminating V and W in favor of the quantities $\rho$ and $A=W/X$, and defining $e^{2h}=X e^{2\nu}$, the spacetime metric can be written in the Papapetrou form 
\begin{eqnarray}\label{Papapetrou}
	ds^2= -\frac{\rho^2}{X(\rho,z)} dt^2 + X(\rho,z)(d\phi + A(\rho,z)dt)^2 + \frac{e^{2h(\rho,z)}}{X(\rho,z)}(d\rho^2 + dz^2).
\end{eqnarray}	 

In the realization of ${\hat {\cal M}}$ as the upper complex half plane, the line segments of $\partial {\hat {\cal M}}$ correspond to the intervals
\begin{eqnarray}
	(-\infty,z_1)(z_1, z_2)(z_2,+\infty),
\end{eqnarray}	 
where the middle interval $(z_1,z_2)$ corresponds to the horizon and the semi-infinite intervals $(-\infty,z_1)$ and $(z_2,+\infty)$ correspond to the southern and northern parts of the rotational axis, respectively. Let us note that the coordinate $z$ is defined in a diffeomorphism invariant way up to shifts by a constant. Therefore, the length of the horizon interval, $l_{H}=z_2-z_1$, is defined invariantly, 
i.e., is the same for any pair of isometric stationary black hole spacetimes of the type we consider. Thus, it may be viewed as a global parameter characterizing the given solution in addition to either the mass $M$ or the angular momentum $J$. By a direct computation, one can prove that 
\begin{eqnarray}
	l_{H}= \frac{1}{2\pi} \kappa {\cal A}_{H} ,
\end{eqnarray}	 
where $\kappa$ ($\kappa>0$) is the surface gravity and ${\cal A}_{H}$ is the area of the horizon. For the Kerr black hole solution, using (\ref{KBHSG}) we find
\begin{eqnarray}
	l_{H}= 2\sqrt{M^2 - \frac{J^2}{M^2}}. 
\end{eqnarray}	 
Inversely, we can express the Kerr black hole mass in terms of the angular momentum $J$ and $l_{H}$,
\begin{eqnarray}\label{HLMR}
	M^2= \frac{1}{8} \left[l^2_{H} + \sqrt{l^4_{H} + 64J^2} \right]
\end{eqnarray}	

{\it We call the collection of all the intervals constituting the boundary $\partial {\hat {\cal M}}$ and the real number $l_{H}$  the  ``interval structure'' of the spacetime.} In fact, the interval structure (in four dimensions) is fully specified by the horizon interval length $l_{H}$, since $z$ is defined up to a constant and the intervals $(-\infty, z_1)$ and $(z_2, +\infty)$ are universal for all stationary, axisymmetric and asymptotically flat spacetimes. In 4-dimensional spacetime the interval structure is more or less trivial, however in higher dimensional spacetimes the interval structure is much more complicated, includes additional topological data and classifies all the stationary and axisymmetric black holes. We refer the interested reader to \cite{Hollands:2007aj, Hollands:2008fm}. 

Having at hand the notion of interval structure we proceed with the proof of the uniqueness theorem. Of key importance is one to carefully monitor the
boundary conditions satisfied by the fields on the boundary of the factor space ${\hat {\cal M}}$. Due to this reason, the dimensionally reduced Einstein 
equations have to be presented in terms of functions which have clear and controllable behavior on $\partial {\hat {\cal M}}$. One such function is the norm $X$ of the Killing field $\eta$ which vanishes only on the axis, i.e. on the intervals $(-\infty,z_1)$ and $(z_2,+\infty)$. The other suitable function which exhibits a good behavior on $\partial {\hat {\cal M}}$ is the twist potential $\chi$ associated with the axial Killing field $\eta$. This potential is introduced in the following way.
First we define the twist $\omega_{\mu}$ of the Killing vector $\omega_{\mu}$ of $\eta$, namely 
\begin{eqnarray}
	\omega_{\mu}=\frac{1}{2}\epsilon^{\mu\nu\alpha\beta}\eta_{\nu}\nabla_{\alpha}\eta_{\beta}.
\end{eqnarray}	
A little bit lengthy but straightforward calculation shows that 
\begin{eqnarray}
	\nabla_{[\mu}\omega_{\nu]}=\frac{1}{2}\epsilon_{\mu\nu\alpha\beta} \eta^{\alpha} R^{\beta\sigma}\eta_{\sigma},
\end{eqnarray} 
which in vacuum gives $\nabla_{[\mu}\omega_{\nu]}=0$. Hence, taking into account that the domain of outer communications is simply connected,
we can conclude that there exists a globally defined potential $\chi$, up to a additive constant, such that 
\begin{eqnarray}
	\omega_{\mu}=\nabla_{\mu}\chi.
\end{eqnarray} 
It is not difficult to show that $\chi$ is invariant under the flow of the Killing fields and therefore is naturally defined scalar field on ${\hat {\cal M}}$ (i.e. $\chi$ depends on $\rho$ and $z$ only). 

On the axis of the rotational symmetry $\eta$ vanishes and therefore $\omega_{\mu}$ vanishes too. Therefore, we have $\chi=const$ on the axis of $\eta$, i.e. 
on the intervals $(-\infty,z_1)$ and $(z_2,\infty)$. The constant value of the potential $\chi$ on the axis intervals we denote by $\chi_{-}$ and ${\chi_{+}}$, respectively. These values are in general different. Since the potential $\chi$ is defined up to an additive constant, we can choose $\chi_{-}=-\chi_{+}$.
The constant $\chi_{+}$ can be determined by calculating the Komar integral for the angular momentum of the black hole
\begin{eqnarray}
	J=\frac{1}{16\pi}\int_{{\cal H}} \nabla_{[\mu}\eta_{\nu]} dS^{\mu\nu} = \frac{1}{8}\int^{z_2}_{z_1}d\chi= \frac{1}{8} (\chi(z_2)-\chi(z_1))=\frac{1}{4}\chi_{+}, 
\end{eqnarray} 
which gives $\chi_{+}=-\chi_{-}=4J$. Since the first derivatives of $\chi$ (i.e. $\omega_{i}={\hat D}_i\chi$) vanish on the axis, near the axis we have 
\begin{eqnarray}\label{NABTP}
	\chi=\pm 4J + O(\rho^2).
\end{eqnarray}

In terms of $X$ and $\chi$ the dimensionally reduced Einstein equations take the form 
\begin{eqnarray}\label{DREFXCHI}
	&&\rho^{-1} \partial_{\rho}\left(\rho \partial_{\rho}X\right)
	+ \partial^2_{z} X = {1\over X} \left[\left(\partial_{\rho} X\right)^2 + \left(\partial_{z} X\right)^2 -
	4 \left(\partial_\rho\chi\right)^2 - 4 \left(\partial_z\chi\right)^2 \right],
	\nonumber \\ 
	&&\rho^{-1} \partial_{\rho}\left(\rho \partial_\rho\chi\right) +
	\partial^2_{z} \chi= {2\over X}\left[\partial_\rho\chi\partial_{\rho} X +
	\partial_z\chi \partial_{z} X \right] ,
\end{eqnarray}
together with the equations for $h$	
\begin{eqnarray}\label{DREFH}
	&&\rho^{-1}\partial_{\rho}h= {1\over 4X^2}\left[(\partial_{\rho}X)^2
	- (\partial_{z}X)^2 + 4 (\partial_\rho\chi)^2 - 4 (\partial_z\chi)^2\right], \nonumber \\
	&&\rho^{-1}\partial_{z}h = {1\over 2X^2}\left[\partial_{\rho}X\partial_{z}X +
	4 \partial_\rho\chi \partial_z\chi\right], \\ 
	&& \partial^2_\rho h + \partial^2_z h= - \frac{1}{4X^2} \left[(\partial_\rho X)^2 + (\partial_z X)^2 + 4 (\partial_\rho \chi)^2 + 4(\partial_z \chi)^2 \right]\nonumber 
\end{eqnarray}
and for the metric function $A$ 	 
\begin{eqnarray}\label{DREFA}
	&&\rho^{-1}\partial_{\rho}A = {2\over X^2} \partial_z\chi , \\ 
	&&\rho^{-1}\partial_{z}A = - {2\over X^2} \partial_\rho \chi, \nonumber
\end{eqnarray}

It is clearly seen that the equations (\ref{DREFXCHI}) for $X$ and $\chi$ are decoupled from the system of equations (\ref{DREFH}) for $h$ and the system (\ref{DREFA}) for the metric function $A$. Once the solution of the system (\ref{DREFXCHI}) is known, we can determine the metric functions $h$ and $A$. Therefore, the problem of the classification of the stationary Einstein black holes can be studied as a 2-dimensional boundary value problem for the nonlinear partially differential equation system (\ref{DREFXCHI}), where the boundary conditions are specified in more detail below.

We have now everything necessary to complete the proof of the uniqueness theorem for the Kerr black hole. We shall prove the theorem in a bit modified version, namely:

\medskip
\noindent

{\bf Theorem} {\it Consider two stationary, asymptotically flat, vacuum black hole solutions $({\cal M}, g_{\mu\nu})$ and $({\tilde {\cal M}}, {\tilde g}_{\mu\nu})$. Assume that both solutions have the same values of the angular momentum $J={\tilde J}$ and the same interval structure. Then they are the same (i.e. isometric).
} 

\medskip
\noindent 
The reason for presenting the uniqueness theorem in this modified form is that it is more universal and, as we already mentioned, holds even for higher dimensional vacuum Einstein equations up to the fact that in higher dimensions the definition of the interval structure is much more complicated including addition topological data \cite{Hollands:2007aj, Hollands:2008fm}.

The crucial step in proving the theorem is to properly define a ``difference between two black hole solutions''. This is not straightforward since the system of equations (\ref{DREFXCHI}) is non-linear. The key observation is that the system (\ref{DREFXCHI}) can be written in a matrix form by introducing an appropriate $2\times 2$ matrix ${\mathbb P}$ composed by the functions $X$ and $\chi$. Then, if we have two black hole solutions represented by their matrices ${\mathbb P}$ and ${\tilde {\mathbb P}}$, the  ``difference'' between them is naturally defined by ${\tilde {\mathbb P}}{\mathbb P}^{-1} - {\mathbf{1} }$, where $\mathbf{1}$ is the unit $2\times 2$ matrix.

The very metrix ${\mathbb P}$ is defined by 
\begin{eqnarray}
	{\mathbb P}= \begin{pmatrix} X^{-1} & -X^{-1}\chi \\	-X^{-1}\chi \;\;\; \; & X+ X^{-1}\chi^2 	\end{pmatrix}.
\end{eqnarray}
The matrix ${\mathbb P}$, as we can see, is symmetric, ${\mathbb P}^{T}={\mathbb P},$ and with $\det {\mathbb P}=1$. ${\mathbb P}$ is also positive semi-definite, i.e. it may be written in the form ${\mathbb P}={\mathbb S}^{T}{\mathbb S}$ for some matrix ${\mathbb S}$ of determinant 1.

Then the system (\ref{DREFXCHI}) can be is written as the following equivalent matrix equation 
\begin{eqnarray}\label{MEFVEE}
	{\hat D}_{i}\left(\rho {\mathbb P}^{-1}{\hat D}^{i}{\mathbb P}\right)=0.
\end{eqnarray} 
If we consider two solutions represented by the matrices ${\mathbb P}$ and ${\tilde {\mathbb P}}$, it is possible to combine the divergence matrix identities (\ref{MEFVEE}) for the two solutions into a single identity on the upper complex half plane, called ``Mazur identity'' \cite{Mazur:1982db, Mazur:1984}. It is given by
\begin{eqnarray}\label{MAZURID}
	{\hat D}_{i}(\rho {\hat D}^{i}\sigma) =\rho {\hat g}^{ij} Tr ({\hat {\mathbb N}}_i {\hat {\mathbb N}}_j), 
\end{eqnarray} 
where 
\begin{eqnarray}
	\sigma= Tr({\tilde {\mathbb P}}{\mathbb P}^{-1} - \mathbf{1}), \;\;\; {\hat {\mathbb N}}_i = {\tilde {\mathbb S}}^{-1}({\tilde {\mathbb P}}^{-1}{\hat D}_i{\tilde {\mathbb P}} - {\mathbb P}^{-1}{\hat D}_i {\mathbb P}) {\mathbb S}. 
\end{eqnarray} 
Here in turn ${\mathbb S}$ and ${\tilde {\mathbb S}}$ are matrices such that ${\mathbb P}={\mathbb S}^{T}{\mathbb S}$ and ${\tilde {\mathbb P}}={\tilde {\mathbb S}}^T {\tilde {\mathbb S}}$ hold. The key point about the Mazur identity (\ref{MAZURID}) is that on the left side we have a total divergence, while the term on the right-hand side is non-negative. This structure can be exploited in the following way. 

At this stage it is convenient to view $\sigma$ not as functions on the complex upper half plane but as axially symmetric functions on the standard flat ${\mathbb R}^3 {\tiny }/\{z-axis\}$, by writing points $x=(x_1,x_2,x_3) \in {\mathbb R}^3$ in cylindrical coordinates as $x=(\rho \cos\phi, \rho \sin\phi, z)$. The Mazur identity then gives 
\begin{eqnarray}\label{flat_Poison}
	 \frac{\partial^2 \sigma}{\partial x^2_1 } + \frac{\partial^2 \sigma}{\partial x^2_2 } + \frac{\partial^2 \sigma}{\partial x^2_3 } \ge 0,
\end{eqnarray} 
on ${\mathbb R}^3 {\tiny }/\{z-axis\}$. In addition we have $\sigma\ge 0$, which is easy to see from the explicit form of $\sigma$, namely 
\begin{eqnarray}\label{EFSIGMA}
	\sigma= \frac{({\tilde X} -X)^2 + ({\tilde \chi} - \chi)^2}{{\tilde X} X} . 
\end{eqnarray}

By the maximum principle \cite{Weinstein:1995tg}, if $\sigma$ is globally bounded above on the entire ${\mathbb R}^3$, including the z-axis and infinity, then it vanishes identically. Thus, what remains is to show that 
$\sigma$ is globally bounded. 

Away from the z-axis in ${\mathbb R}^3$ (or equivalently $\partial{\hat M}$) we have $X\ne 0$, ${\tilde X}\ne 0$ and our functions are smooth (in even analytic), and this consequently
means that $\sigma$ is also smooth. The same applies to the horizon interval $(z_1,z_2)$ where ${\tilde X}\ne 0$ and $X\ne 0$, too. 

On the intervals $(-\infty, z_1)$ and $(z_2,+\infty)$ the problem is a bit more tricky because these intervals constitute the rotational axis where $\eta=0$ or equivalently $X=0$ and ${\tilde X}=0$. 
For $X$ and ${\tilde X}$ we have $X=O(\rho^2)$ and ${\tilde X}=O(\rho^2)$ near the axis. According to (\ref{NABTP}), the potentials $\chi$ and ${\tilde \chi}$ behave as $\chi=\pm J + O(\rho^2)$ and ${\tilde \chi}=\pm {\tilde J} + O(\rho^2)$ near the axis of $\eta$. Since $J={\tilde J}$ we find that $\chi-{\tilde \chi}=O(\rho^2)$ near the axis of $\eta$. Therefore, (\ref{EFSIGMA}) shows that $\sigma$ is bounded on the axis of $\eta$.

At the points where the axis of $\eta$ meets the horizon, i.e. the corners $z_1$ and $z_2$, the function $\sigma$ has also to be bounded and this follows from the continuity argument. 

To show that $\sigma$ is bounded near infinity, it is convenient to introduce the asymptotic coordinates ($r$, $\theta$) defined by $\rho=r\sin\theta$ and $z=r\cos\theta$. Straightforward calculations show that, in the asymptotic region $r\to \infty$, $X$ and $\chi$ have the following asymptotic behaviour
\begin{eqnarray}
	&& X=r^2\sin^2\theta (1 + O(1/r^2)), \nonumber \\ && \chi = 2J \cos\theta (2+ \sin^2\theta) + O(1/r^2),
\end{eqnarray} 
which in turn shows that $\sigma= O(1/r^2)$ .

Summarizing, when viewed as an axisymmetric function on ${\mathbb R}^3$, the function $\sigma$ is a solution to eq.(\ref{flat_Poison}) which is bounded on the entire space ${\mathbb R}^3$, including the z-axis. 
Therefore $\sigma$ vanishes globally. In other words, we have $X={\tilde X}$ and $\chi={\tilde \chi}$ which proves the theorem. 

We proved that there is only one black hole solution to the vacuum Einstein equations with a regular horizon and a stationary, asymptotically flat domain of outer communication for given angular momentum $J$ and length $l_h$ of the horizon interval. This is just the Kerr solution with given $J$ and $l_H$, or equivalently the Kerr solutions with given angular momentum $J$ and given mass $M^2= \frac{1}{8} \left[l^2_{H} + \sqrt{l^4_{H} + 64J^2} \right] $ (see eq.(\ref{HLMR})). Since the mass $M$ and the horizon interval length $l_H$ are in one-to-one correspondence for fixed $J$, we have proven that the vacuum stationary black holes in General Relativity are fully specified by their mass $M$ and the angular momentum $J$.

This uniqueness theorem reveals the remarkable simplicity of the vacuum black holes in general relativity -- they are fully specified only by their asymptotic conserved charges, namely the mass and the angular momentum.

\section{Vacuum black holes in classical scalar-tensor theories \\ with a single scalar field}\label{VBHSSF}

We focus on the vacuum sector of the classical scalar-tensor theories. Then, as follows from (\ref{FEAF}), the Einstein frame vacuum field equations are just the Einstein equations with a minimally coupled scalar field possessing a potential, namely 
\begin{eqnarray}\label{VFEAF}
	&&R_{\mu\nu} -\frac{1}{2}R g_{\mu\nu} = 2\nabla_{\mu}\varphi\nabla_{\nu}\varphi - g_{\mu\nu} \nabla_{\sigma}\varphi \nabla^{\sigma}\varphi -	{1\over 2}g_{\mu\nu}V(\varphi) ,\nonumber \\
	&&\nabla_{\mu}\nabla^{\mu}\varphi =	{1\over 4}V^{{\,}\prime}(\varphi). 	
\end{eqnarray} 
In what follows assume that the scalar potential $V(\varphi)$ is non-negative 
\begin{eqnarray}\label{PositivV}
	V(\varphi)\ge 0,
\end{eqnarray}
and that $V^{\prime}(\varphi)=0$ possesses at least one discrete solution $\varphi_{*}$ at which $V(\varphi_*)=0$. Without loss of generality, we shall put $\varphi_*=0$. 

Taking into account that $V(\varphi)\ge 0$ it can be shown that the (Einstein frame) energy-momentum of the scalar field 
\begin{eqnarray}
	T^{\varphi}_{\mu\nu}=2\partial_{\mu}\varphi\partial_{\nu}\varphi -g_{\mu\nu}\nabla_{\alpha}\varphi\nabla^{\alpha}\varphi - {1\over 2}g_{\mu\nu}V(\varphi)
\end{eqnarray} 
satisfies the weak and null energy condition, i.e. for every future-pointing timelike or null vector $u^{\mu}$ we have $T^{\varphi}_{\mu\nu}u^\mu u^\nu\ge 0$. 

We now turn to the question of whether the stationary, vacuum scalar-tensor equations (\ref{VFEAF}) admit black hole solutions with a non-trivial scalar field outside the black hole, i.e. black holes with scalar hair.

Here we shall consider the case when the stationary Killing field $\xi$ is not tangent to the null generators of the horizon\footnote{When $\xi$ is tangent to the null generators of the horizon, the spacetime time must be static. The static case can be treated pretty much in the same way.}. Then, as follows from the rigidity theorem, there exists an additional axial Killing field $\eta$ which commutes with $\xi$ and is such that ${\cal L}_{\eta}\varphi={\cal L}_{\xi}\varphi=0$. In other words, the spacetime has isometry group ${\mathbb R}\times SO(2)$ and the scalar field is also stationary and axisymmetric.
This also means that the gradient of the scalar field must be spacelike or zero everywhere in the domain of outer communications. 

Below we present a non-hair theorem and sketch its proof by following the classical arguments of Hawking \cite{Hawking:1972qk} (see also \cite{Bekenstein:1972ny} and \cite{Sotiriou:2011dz, Herdeiro:2015waa}). 

Let us consider a partial Cauchy surface ${\cal J}$ in the domain of outer communications and another copy ${\cal J}^{\prime}\ne {\cal J} $ of ${\cal J}$ obtained
from ${\cal J}$ under the flow of the Killing field $\xi$. Let ${\cal D}$ be the region bounded by ${\cal J}$, ${\cal J}^{\prime }$, a portion of the event horizon
$H$ and a timelike 3-surface ${\cal T}$ at infinity.

As a next step, we consider the equation for the scalar field and multiply it by $\varphi$ which, after some manipulation, gives
\begin{eqnarray}\label{MKGI}
	\nabla_{\mu}(\varphi\nabla^{\mu}\varphi)= \nabla_{\mu}\varphi\nabla^{\mu}\varphi + \frac{1}{4}\varphi V^{\prime}(\varphi).
\end{eqnarray}
Integrating (\ref{MKGI}) over ${\cal D}$ and using the Gauss theorem, we can express the left-hand side as surface integral on the boundary $\partial {\cal D}$ of ${\cal D}$, namely 
\begin{eqnarray}\label{MKGI2}
	\int_{\partial{\cal D}} \!\! d^3S n_{\mu}(\varphi\nabla^{\mu}\varphi)\! =\!\!\int_{{\cal D}}\!\! d^4x\sqrt{-g} \left[\nabla_{\mu}\varphi\nabla^{\mu}\varphi + \frac{1}{4}\varphi V^{\prime}(\varphi)\right]\!,
\end{eqnarray}
where $n_{\mu}$ is the unit normal to $\partial{{\cal D}}$. Now, taking into account that the structure of the boundary ${\partial {\cal D}}$, i.e. ${\partial {\cal D}}= {\cal J} \cup {\cal J}^{\prime} \cup {\cal T} \cup H$ we get 
\begin{eqnarray}
	&&	\int_{\partial{\cal D}} d^3S n_{\mu}(\varphi\nabla^{\mu}\varphi)= \int_{{{\cal J}}} d^3S n_{\mu}(\varphi\nabla^{\mu}\varphi) + \int_{{{\cal J}^\prime}} d^3S n_{\mu}(\varphi\nabla^{\mu}\varphi) \notag\\ 
	&&+ \int_{{\cal T}} d^3S n_{\mu}(\varphi\nabla^{\mu}\varphi)+ \int_{{\cal H}} d^3S n_{\mu}(\varphi\nabla^{\mu}\varphi).
\end{eqnarray}
 ${\cal J}$ and ${\cal J}^{\prime}$ are isometric copies and have opposite directions. Therefore the surface integral over ${\cal J}$ cancels out that over ${\cal J}^{\prime}$. The black hole horizon integral gives no contribution because on the horizon $n^{\mu}=K^{\mu}=\xi^\mu + \Omega_H \eta^\mu$ and $n_{\mu}\nabla^{\mu}\varphi=K_{\mu}\nabla^{\mu}\varphi =\xi_{\mu}\nabla^{\mu}\varphi+ \Omega_{H}\eta_{\mu}\nabla^{\mu}\varphi=0$. The surface integral at infinity is also zero because $\varphi$ is zero there -- the asymptotic flatness requires $\varphi_{\infty}=\varphi_*=0$. In this way we find that 
\begin{eqnarray}\label{IFZSF}
	\int_{{\cal D}}d^4x\sqrt{-g} \left[\nabla_{\mu}\varphi\nabla^{\mu}\varphi + \frac{1}{4}\varphi V^{\prime}(\varphi)\right]=0.
\end{eqnarray}
If we assume that $\varphi V^{\prime}(\varphi)\ge 0$ everywhere and taking into account that the scalar field gradient is spacelike or zero, i.e. $\nabla_{\mu}\varphi\nabla^{\mu}\varphi\ge 0$, we conclude that (\ref{IFZSF}) is satisfied if and only if $\nabla_{\mu}\varphi\nabla^{\mu}\varphi=0$ and 
$\varphi V^{\prime}(\varphi)=0$. This in turn gives $\varphi=0$ everywhere in the domain of outer communications. In this case the equations (\ref{VFEAF}) are the same as the Einstein equations. Thus, stationary black hole solutions in vacuum scalar-tensor theories with a single field are the same as the stationary black hole solutions in the Einstein theory and they are completely represented by the Kerr family of solutions.

A similar result can be derived if we replace the condition $\varphi V^{\prime}(\varphi)\ge 0$ with the condition $V^{\prime\prime}(\varphi)\ge 0$ when
the potential is not identically zero. Following a very similar procedure, we multiply the equation for the scalar field by $V^{\prime}(\varphi)$ and after some algebra we find 
\begin{eqnarray}\label{MKGI1}
	\nabla_{\mu}(V^{\prime}(\varphi)\nabla^{\mu}\varphi)= V^{\prime\prime}(\varphi)\nabla_{\mu}\varphi\nabla^{\mu}\varphi + \frac{1}{4} (V^{\prime}(\varphi))^2.
\end{eqnarray}
Integrating over the region ${\cal D}$ and applying the Gauss theorem, we get
\begin{eqnarray}\label{PI1}
	\int_{\partial{\cal D}} d^3S n_{\mu}(V^{\prime}(\varphi)\nabla^{\mu}\varphi) 
	=\int_{{\cal D}}d^4x\sqrt{-g} \left[ V^{\prime\prime}(\varphi)\nabla_{\mu}\varphi\nabla^{\mu}\varphi + \frac{1}{4} (V^{\prime}(\varphi)^2\right].
\end{eqnarray}
The same arguments as in the previous case show that the surface integral over ${\partial\cal D}$ vanishes and therefore the right-hand side of (\ref{PI1}) is zero which means
that $\nabla_{\mu}\varphi\nabla^{\mu}\varphi=0$ and $V^{\prime}(\varphi)=0$ everywhere. Hence we can conclude that $\varphi=0$ everywhere 
in the domain of outer communications. 

Summarizing the results so far we have the following no-scalar hair theorem: 

\medskip
\noindent

{\bf Theorem} {\it Let us consider the vacuum equations of the classical scalar-tensor theories with a non-negative scalar potential $V(\varphi)$ satisfying the condition $\varphi V^{\prime}(\varphi)\ge 0$ or $V^{\prime\prime}(\varphi)\ge 0$ everywhere. Then the only black hole solutions to these equations with regular event horizon and stationary and asymptotically flat domain of outer communications is the Kerr family of solutions. } 

\medskip
\noindent

The additional conditions imposed on the scalar potential lead to some shortcomings of the above no-hair theorem. For example, the theorem does not cover Higgs-like potentials. What is desirable from a physical point of view is the non-negativeness of the potential to be the only condition imposed on it. Up to now, no such general result has been proven. However, if in addition a spherical symmetry is assumed, the extra conditions on the potential can be relaxed at that price \cite{Heusler:1992ss},\cite{Bekenstein:1995un},\cite{Sudarsky:1995zg}. The proof in this case will be given in the next sections in the more general setting of multiple scalar fields with an arbitrary target space.

In any case, the above theorem imposes strong restrictions on the possible black hole solutions in the classical scalar-tensor theories with a single scalar field -- 
the stationary black hole solutions are the same as those in General Relativity. In simple words, the scalar field must be trivial and the geometry
must be described by the Kerr metric.

It is worth mentioning, however, that, in the presence of matter, the black holes in scalar-tensor theories with a single scalar field can in general have an ``indirect'' scalar hair sourced by the matter surrounding the black holes \cite{Cardoso:2013fwa}. This in turn changes the astrophysics around the black holes in comparison to that in GR.

\section{Black holes in multiscalar theories of gravity}\label{SEC_MSBH}

Contrary to common expectations, the qualitative picture of the black hole solutions within the classical multiscalar theories of gravity is rather different from that in the scalar-tensor theories with a single scalar field. In contrast to the case with a single scalar field, in the presence of multiple massive scalar fields, black hole solutions different from the Kerr solution can exist \cite{Herdeiro:2014goa, Collodel:2020gyp}. The simple reason behind this is the fact that in the multiscalar theories, the scalar fields can be dynamical, i.e. time-dependent (time-periodic), while leaving the spacetime geometry stationary. The dynamics of the scalar fields can be confined in the target space of scalar fields in such a way that the energy-momentum tensor of the scalar fields is time-independent. This occurs when the black holes are rotating and the massive scalar fields depend on the time and azimuthal coordinate so that the scalar fields flux through the horizon can be cancelled. 

There are no general theorems classifying the black hole solutions in the multiscalar theories of gravity. Even more, the rotating black holes in these theories are constructed only numerically. That is why, taking the complicated character of the black hole solution spectrum in multiscalar theories, we shall divide this section into subsections for a clearer presentation and to demonstrate various mathematical techniques for proving no-hair theorems.

\subsection{Static black holes in the vacuum multiscalar theories with $V(\varphi)=0$ and static scalar fields}
In this section, we shall focus on the vacuum multiscalar theories of gravity with a trivial scalar potential, $V(\varphi)=0$. Then the vacuum field equations
from section (\ref{Multi}) can be written in the form 
\begin{eqnarray}\label{VMSGFE}
	&&R_{\mu\nu} =
	2{\gamma}_{ab}(\varphi)
	\nabla_{\mu}\varphi^{a}\nabla_{\nu}\varphi^{b},
	\\ 
	&&\nabla_{\mu}\nabla^{\mu}\varphi^{a}
	+ {\gamma^{\,a}}_{bc}(\varphi)
	\nabla_{\mu}\varphi^{b}\nabla^{\mu}\varphi^{c}=0. \nonumber
\end{eqnarray}
We consider static, asymptotically flat black hole spacetimes satisfying field the equations (\ref{VMSGFE}) with the assumption that the scalar fields $\varphi^a$ are also static, i.e. they satisfy 
\begin{eqnarray}
	{\cal L}_{\xi}\varphi^a =0,
\end{eqnarray}
where $\xi$ is the (hypersurface orthogonal) stationary Killing vector. It is worth noting that, to some extent, this condition is restrictive since, as we shall see in the other sections, the scalar fields can be time-dependent while their energy-momentum tensor is still stationary. 

The spacetime metric is given by (\ref{SMETRIC}) and the dimensionally reduced equations take the form 
\begin{eqnarray}\label{DRSFEMSTT}
	&&R^{(3)}_{ij}=N^{-1}D_{i}D_{j}N + 2\gamma_{ab}(\varphi)D_{i}\varphi^a D_{j}\varphi^b, \notag \\
	&&D^{i}D_{i}N=0,\\
	&&D_{i}D^{i}\varphi^a + N^{-1}D_{i}ND^{i}\varphi^a + \gamma^{a}_{bc}(\varphi) D_{i}\varphi^b D^{i}\varphi^c = 0, \notag
\end{eqnarray}
where $D_i$ and $R^{(3)}_{ij}$ are Levi-Civita connection and the Ricci tensor associated with the Riemannian manifold $(\Sigma,g^{(3)}_{ij})$ defined in section \ref{SBHGR}. By contracting the first of equations (\ref{DRSFEMSTT}) we also get 
\begin{eqnarray}\label{DRRSMST}
	R^{(3)}= 2\gamma_{ab}(\varphi)D_{i}\varphi^a D^{i}\varphi^b .
\end{eqnarray}

We shall prove the following theorem \cite{Heusler:1993cd}

\medskip
\noindent

{\bf Theorem }{\it The vacuum field equations of the multiscalar theories with $V(\varphi)=0$ and static scalar fields, admit only the Schwarzschild solution as black hole solution with regular horizon and static, asymptotically flat domain of outer communications.} 

\medskip
\noindent

The proof is pretty much the same as in the vacuum case discussed in section (\ref{SBHGR}) and that is why we shall only
outline the main steps. We consider again two copies $\Sigma_{+}$ and $\Sigma_{-}$ of $\Sigma$, with metrics $g^{(3\pm)}_{ij}=\Omega^2_{\pm} g^{(3)}_{ij}$ where 
the conformal factors $\Omega_{\pm}$ are the same as in the general relativistic case, i.e. they are given by (\ref{CFGR}). In the present case the function $N$ is also harmonic and the same reasoning as in section (\ref{SBHGR}) shows that the conformal transformations are regular. The manifold ${\hat \Sigma_{+\cup -}}$ obtained by passting the two copies $\Sigma_{+}$ and $\Sigma_{-}$ along the common boundary has a metric ${\hat g}^{(3)}_{ij}$ continuous accross the horizon cross section ${\cal H}$ since $N=0$ there. The second fundamental form also matches continuously on ${\cal H}$ which guarantees the continuity of the Ricci tensor ${\hat R}_{ij}$. The same arguments as in the pure vacuum case show that the infinity related to $\Sigma_{-}$ can be compactified by adding the infinite point. The so-constructed space ${\hat \Sigma}={\Sigma}_{+}\cup \Sigma_{-}\cup \infty$ is complete with only one asymptotic region and direct calculation shows that its mass ${\hat M}$ vanishes. 
To conclude the proof, we have to show that the Ricci scalar ${\hat R}$ is non-negative. With the help of (\ref{CT}) we find 
\begin{eqnarray}\label{DRRSMSTT}
	\Omega^4{\hat R}^{(3)} = \Omega^2_{\pm} R^{(3)} \mp 2\Omega_{\pm}(1 \pm N)\Omega D_{i}D^{i}N = 
	2\Omega^2_{\pm}\gamma_{ab}(\varphi)D_{i}\varphi^a D^{i}\varphi^b,
\end{eqnarray} 
where we have taken into account that $D_{i}D^{i}N =0$ and (\ref{DRRSMST}). Now it is clear that $R^{(3)}$ is non-negative. 

We have that the Riemannian manifold $({\hat \Sigma}, {\hat g}^{(3)}_{ij})$ satisfies the conditions required for the application of the positive
mass theorem (see section (\ref{SBHGR})), from which we now conclude that $({\hat \Sigma}, {\hat g}^{(3)}_{ij})$ is isometric to
$({\mathbb R}^3, \delta_{ij})$. Equation (\ref{DRRSMSTT}) now implies that the scalar fields have to be constant and therefore the case reduces to the vacuum one.
This completes the proof.

\subsection{Static spherically symmetric black holes in multiscalar theories \\ with non-negative potential $V(\varphi)$ and static scalar fields}

As we already mentioned, general black hole no-hair theorems within the multiscalar theories with non-negative potential and with no further conditions on it and on the target space, have been proven only under the additional assumption of spherical symmetry \cite{Heusler:1992ss, Doneva:2020dji}. That is why we focus first on the static and spherically symmetric problem. The Einstein frame field equations are the following (see section (\ref{Multi})):
\begin{eqnarray}\label{FE1}
&&R_{\mu\nu} - \frac{1}{2}R g_{\mu\nu} = 2\gamma_{ab}(\varphi)\nabla_{\mu}\varphi^a \nabla_{\nu}\varphi^b 
	- g_{\mu\nu}\gamma_{ab}\nabla_{\sigma}\varphi^a\nabla^{\sigma}\varphi^b
	- \frac{1}{2} V(\varphi) g_{\mu\nu}, \\
&&\nabla_{\mu}\nabla^{\mu}\varphi^a= 
	- \gamma^{a}_{cd}(\varphi)\nabla_{\mu}\varphi^c \nabla^{\mu}\varphi^d + \frac{1}{4}\gamma^{ab}(\varphi)\frac{\partial V(\varphi)}{\partial\varphi^b}. \notag 
\end{eqnarray}

The spacetime metric takes the well-known form 
\begin{eqnarray}\label{SSM}
	ds^2= - e^{2\Phi(r)}dt^2 + e^{2\Lambda(r)}dr^2 + r^2 s_{ij}dx^idx^j,
\end{eqnarray} 
where $s_{ij}$ is the metric on the unit 2D sphere, namely $s_{ij}dx^idx^j=d\theta^2 + \sin^2\theta d\phi^2$.

We shall consider static scalar fields, ${\cal L}_{\xi}\varphi^a=0$, where $\xi=\frac{\partial}{\partial t}$. This automatically ensures that the effective energy-momentum tensor of the scalar fields is static, ${\cal L}_{\xi} T^{\varphi}_{\mu\nu}=0$. Contrary to the static symmetry, the scalar fields are not required to respect the spherical symmetry, i.e. to depend only on the radial coordinate $r$. They can depend on the angular variables $\theta$ and $\phi$ in such a way that the energy-momentum tensor respects the spherical symmetry. 

Under the assumptions we made, the field equations reduce to the following system 
\begin{eqnarray}\label{DRE0} 
	&&\frac{2}{r}e^{-2\Lambda} \Lambda^{\prime} + \frac{1}{r^2}\left(1-e^{-2\Lambda}\right)= \frac{1}{2}K + \frac{1}{2}V(\varphi), \\ 
	\notag \\
	&&\frac{2}{r}e^{-2\Lambda} \Phi^{\prime} - \frac{1}{r^2}\left(1-e^{-2\Lambda}\right)\!= 
	e^{-2\Lambda}\gamma_{ab}(\varphi)\partial_r\varphi^a \partial_r\varphi^b - \frac{1}{r^2}s^{ij}\gamma_{ab}\partial_{i}\varphi^a\partial_{j}\varphi^b - \frac{1}{2}V(\varphi), \\ 
	\notag \\
	&&e^{-2\Lambda}\left[\Phi^{\prime\prime} + (\Phi^{\prime} + \frac{1}{r})(\Phi^{\prime} - \Lambda^{\prime})\right]\! r^{2} s_{ij}=
	2 \gamma_{ab}(\varphi)\partial_i\varphi^a \partial_j\varphi^b - \frac{1}{2} r^2 s_{ij} \left[K + V(\varphi)\right], 
\end{eqnarray} 
\begin{eqnarray}\label{EQFF0} 	
	\partial_r \left(e^{\Phi-\Lambda} r^2 \gamma_{ab}(\varphi)\partial_r\varphi^b \right) +
	\frac{e^{\Phi + \Lambda}}{\sqrt{s}} \partial_i\left(\sqrt{s} s^{ij} \gamma_{ab}(\varphi)\partial_j\varphi^b\right) = \frac{e^{\Phi + \Lambda}}{4} r^2 \frac{\partial K}{\partial \varphi^a}.
\end{eqnarray} 
where 
\begin{eqnarray}
	K=2\gamma_{ab}\nabla_{\mu}\varphi^a \nabla^{\mu}\varphi^b= 2e^{-2\Lambda}\gamma_{ab}(\varphi)\partial_r\varphi^a\partial_r\varphi^b
	 + \frac{2}{r^2}s^{ij}\gamma_{ab}(\varphi)\partial_{i}\varphi^a\partial_{j}\varphi^b . 
\end{eqnarray}

Due to the spherical symmetry and using the dimensionally reduced field equations, it follows that $V(\varphi)$, $\gamma_{ab}(\varphi)\partial_r\varphi^a \partial_r\varphi^b $ and $s^{ij}\gamma_{ab}(\varphi)\partial_i\varphi^a \partial_j\varphi^b$ are functions of $r$ only. We can then define
\begin{eqnarray}
	P^2(r)=\gamma_{ab}(\varphi)\partial_r\varphi^a \partial_r\varphi^b, \; \; \; \;
	H^{2}(r)= s^{ij}\gamma_{ab}(\varphi)\partial_i\varphi^a \partial_j\varphi^b .
\end{eqnarray}

For asymptotically flat spacetimes, using the dimensionally 
reduced field equations, one can show that 
\begin{eqnarray}
	\lim_{r\to \infty} V(\varphi)=0, \;\;\; \lim_{r\to \infty} P^2(r)=0\;\;\; \lim_{r\to \infty} H^2(r)=0.
\end{eqnarray} 
More precisely $V(\varphi)$, $P^2(r)$ and $H^2(r)$ drop off at least as 
\begin{eqnarray}
	V(\varphi) \sim \frac{1}{r^4}, \; \; \;\;\;\; P^2(r) \sim \frac{1}{r^4} , \;\;\;\;\; \;\; H^2(r)\sim \frac{1}{r^2}
\end{eqnarray}
for $r\to \infty$. 

We proceed with the derivation of a divergence identity which plays a key role in proving the no-hair theorem. The potential application of the divergence identity is, of course, beyond the no-hair theorems. It can also be used to study the quantitative and qualitative properties of static and spherically symmetric solutions in theories with multiple scalar fields.


The divergence identity can be derived by using the conservation of the scalar fields energy-momentum tensor, namely $ \nabla_{\mu}T^{\mu}_{\nu}=0$. This equation written in explicit form is
\begin{eqnarray}
	e^{-(\Phi + \Lambda)} \frac{d}{dr}\left[e^{\Phi + \Lambda} T^{r}_{r} \right] = \frac{1}{2r}e^{2\Lambda}\left[(1-e^{-2\Lambda}) (T^{t}_{t} - T^{r}_{r}) 
	+ 2e^{-2\Lambda}(2T^{\theta}_{\theta} - T^{t}_{t} - T^{r}_{r`})\right], \notag
\end{eqnarray}
where 
\begin{eqnarray}
	&&T^{t}_{t} = -e^{-2\Lambda}P^2(r) - \frac{H^{2}(r)}{r^2} - \frac{1}{2}V(\varphi), 	\\
	&&T^{r}_{r} = e^{-2\Lambda}P^2(r) - \frac{H^{2}(r)}{r^2}- \frac{1}{2}V(\varphi), \\
	&& T^{\theta}_{\theta}=T^{\phi}_{\phi}= -e^{-2\Lambda}P^2(r) - \frac{1}{2}V(\varphi).
\end{eqnarray} 
Substituting back these expressions for the components of the energy-momentum tensor and after some algebra we find 
\begin{eqnarray}
	\frac{d}{dr}\left[e^{\Phi + \Lambda} \left(\frac{H^{2}(r)}{r^2}+ \frac{1}{2}V(\varphi) - P^2(r)e^{-2\Lambda} \right)\right] 
	=\frac{e^{\Phi + \Lambda}}{r}\left[(1+ 3e^{-2\Lambda})P^2(r) - 2 \frac{H^2(r)}{r^2}\right] \notag .
\end{eqnarray} 
So obtained divergence identity, however, is not what we want since its right-hand side is not manifestly non-negative. To obtain a divergence identity with a manifestly non-negative right-hand side, we write the identity in the form 
\begin{eqnarray}
	\frac{dB}{dr}= A.
\end{eqnarray}
Then, for a function $\Omega(r)$, we can construct a new identity of the form
\begin{eqnarray}
	\frac{d (\Omega B)}{dr}= \Omega \left(A + \frac{\Omega^\prime}{\Omega}B\right). 
\end{eqnarray} 
Finally, choosing $\Omega(r)=r^2$ we get the desired divergence identity
\begin{eqnarray}\label{DISC}
	\frac{d}{dr}\left[e^{\Phi+\Lambda} H^2(r) + \frac{1}{2}e^{\Phi+\Lambda}r^2 V(\varphi) -
	e^{\Phi-\Lambda} r^2 P^2(r) \!\right] 
	\!\!=\! r e^{\Phi + \Lambda} \!\left[(1+ e^{-2\Lambda}) P^2(r) + V(\varphi) \right]. 
\end{eqnarray}


Having at hand the desired divergence identity we are ready to prove our no-hair theorem. 
Recall that we consider theories with $V(\varphi)\ge 0$. When this condition is satisfied, the right-hand side of the divergence identity is non-negative. 
Integrating then the divergence identity from the horizon $r=r_h$ to infinity we get 
\begin{eqnarray}\label{BHInt0}
	- e^{(\Phi+\Lambda)_h} \left[H^2(r_h) + \frac{1}{2}r_h^2 V(\varphi_h)\right] 
	= \int^{\infty}_{r_h} dr r e^{\Phi + \Lambda} \left[(1+ e^{-2\Lambda}) P^2(r) + V(\varphi) \right], 
\end{eqnarray}
where we have taken into account that $\lim_{r\to \infty} H^2(r)=0$, $\lim_{r\to \infty} r^2 P^2(r)=0$, $\lim_{r\to \infty} r^2V(\varphi)=0$ as well as $(e^{\Phi-\Lambda} r^2 P^2(r))_h=0$. Note that $(\Phi+\Lambda)_h$ is finite for regular horizons.
The right-hand side of (\ref{BHInt0}) is non-negative while the left-hand side is non-positive. Therefore we can conclude that
both sides vanish. Consequently we have $P^2(r)=0$ and $V(\varphi)=0$ for every $r\in [r_h,\infty)$ as well as 
$H^2(r_h)=0$. Substituting $P^2(r)=0$ and $V(\varphi)=0$ in the divergence identity (\ref{DISC}) we find that $H^2(r)=0$ for every $r\in [r_h,\infty)$. As a consequence of all this, we conclude that the map $\varphi$ is a constant map $\varphi=\varphi_0$ with $V(\varphi_0)=0$. 

All the results above show that the right-hand side of the dimensionally reduced field equations (\ref{DRE0})-(\ref{EQFF0}) vanish and the equations are reduced to the vacuum, static, and spherically symmetric Einstein equations whose unique black hole solution with a regular horizon is the Schwarzschild one. Summarizing, we proved the following

\medskip
\noindent

{\bf Theorem } {\it Let us consider the vacuum equations of the classical multiscalar theories of gravity and assume that the scalar field potential is non-negative, $V(\varphi)\ge 0$. 	Then every static and spherically symmetric black hole solution to the field equations (\ref{FE1}) with static multiscalar map $\varphi$ (${\cal L}_\xi\varphi$=0) and regular horizon consists of the Schwarzschild solution and a constant map $\varphi_0$ with $V(\varphi_0)=0$. } 
\medskip
\noindent

It is instructive to present yet another proof of the no-hair theorem based on simple and beautiful scaling arguments \cite{Heusler:1992ss}. For this purpose let us consider the local mass $m(r)=\frac{r}{2} (1-e^{-2\Lambda(r)})$ which, as follows from the equation for $\Lambda$, satisfies the equation
\begin{eqnarray}
	m^\prime (r) = \frac{1}{2}e^{-2\Lambda(r)} r^2 P^2(r) + \frac{1}{2}H^2(r) + \frac{1}{4} r^2V(\varphi). 
\end{eqnarray}
By substituting $e^{-2\Lambda(r)}= 1 - 2m(r)/r$ in the first term on the right side of the above equation we get 
\begin{eqnarray}
	m^\prime (r) = \frac{1}{2} r^2 P^2(r) + \frac{1}{2}H^2(r) + \frac{1}{4} r^2V(\varphi) - m(r) r P^2(r). 
\end{eqnarray}
From the dimensionally reduced equations, we have $r P^2(r)= (\Phi(r) + \Lambda(r))^\prime=-\delta^\prime(r) $ which, substituted in the equation for the local mass, leads to 
\begin{eqnarray}
	m^\prime (r) = \frac{1}{2} r^2 P^2(r) + \frac{1}{2}H^2(r) + \frac{1}{4} r^2V(\varphi) + m(r) \delta^\prime(r) . 
\end{eqnarray}

Now let us define the local mass $\mu(r)$ of the scalar hair outside the black hole horizon, namely $\mu(r)= m(r) - m(r_h)= m(r) - \frac{r_h}{2} $. Using the equation for $m(r)$, one can show that
\begin{eqnarray}
	\mu^\prime (r) = \frac{1}{2} r^2 P^2(r) + \frac{1}{2}H^2(r) + \frac{1}{4} r^2V(\varphi) + \mu(r)\delta^\prime(r) + \frac{r_h}{2} \delta^\prime(r), 
\end{eqnarray}
which can also be rewritten as
\begin{eqnarray}
	\mu^\prime (r) = \frac{1}{2} (1 - \frac{r_h}{r}) r^2 P^2(r) + \frac{1}{2}H^2(r) + \frac{1}{4} r^2V(\varphi) + \mu(r) \delta^\prime(r), 
\end{eqnarray}
or equivalently in the form
\begin{eqnarray}
	\left( \mu(r) e^{-\delta(r)}\right)^\prime = \left[\frac{1}{2} (1 - \frac{r_h}{r}) r^2 P^2(r) + \frac{1}{2}H^2(r) + \frac{1}{4} r^2V(\varphi)\right] e^{-\delta(r)}. 
\end{eqnarray} 

The total mass $\mu(\infty)$ of the scalar hair outside the horizon can then be found by integrating the above identity from the horizon to infinity which gives
\begin{eqnarray}
	\mu(\infty) = \int^{\infty}_{r_h} \left[\frac{1}{2} (1 - \frac{r_h}{r}) r^2 P^2(r) + \frac{1}{2}H^2(r) + \frac{1}{4} r^2V(\varphi)\right] e^{-\delta(r)} dr, 
\end{eqnarray}
 where we have taken into account that $\mu(r_h)=0$ and $\delta(\infty)=0$.

For applying the scaling arguments we need the dimensionless coordinate $x=r/r_h$. Then, in terms of $x$ we have 
\begin{eqnarray}	
	\mu(\infty) = r_h\int^{\infty}_{1} \left[\frac{1}{2} (1 - \frac{1}{x}) x^2 P^2(x) + \frac{1}{2}H^2(x) + \frac{1}{4} r_h^2 x^2 V(\varphi)\right] e^{-\delta(x)} dx. 
\end{eqnarray}

Next, we consider the total mass of the scalar hair for a one-parameter family of scalar fields $\varphi^a(\lambda,x, \theta,\phi)$ where $\varphi^a(\lambda=1,x, \theta,\phi)$ is a solution to our system of equations.
In order for the $\lambda$-deformations to vanish on the horizon we chose $\varphi^a(\lambda,x, \theta,\phi)=\varphi^a(x^\lambda, \theta,\phi)$. In calculating $\mu(\infty,\lambda)$ we perform the coordinate change $y=x^{\lambda}$ and we get
\begin{eqnarray}	
	\mu(\infty,\lambda) = r_h\int^{\infty}_{1} \left[\frac{\lambda}{2} (y^\frac{1}{\lambda}-1) y P^2(y) + \frac{1}{2\lambda} y^{\frac{1}{\lambda}-1} H^2(y) + \frac{1}{4\lambda} r_h^2 y^{\frac{3}{\lambda} -1} V(\varphi)\right] e^{-\lambda\delta(y)} dy. 
\end{eqnarray}
Here we have also taken into account that $\delta(\lambda,y)=\lambda \delta(y)$ which can be derived from the equation $r P^2(r)=-\delta^\prime(r) $.

In order for $\varphi^a(\lambda=1,x, \theta,\phi)$ to be solution we must have $\frac{d\mu(\lambda,\infty)}{d\lambda} (\lambda=1)=0$ which gives

\begin{eqnarray}\label{SCALINGCON}	
 \int^{\infty}_{1} \left[\frac{1}{2} [(1-\frac{1}{y}) \delta(y) +\sigma(y) ] y^2 P^2(y) + \frac{1}{2} [1 + \ln y + \delta(y)] H^2(y) \nonumber \right. \\ + \left. \frac{1}{4} [1 + 3\ln y + \delta(y) ] r_h^2 y^{2} V(\varphi)\right] e^{-\delta(y)} dy = 0, 
\end{eqnarray}
where $\sigma(y)= \frac{1}{y} +\ln y - 1$. It is not difficult one to see that $\sigma(y)\ge 0$ for $y\ge 1$. In addition we have $\delta(y)\ge 0$ since 
\begin{eqnarray}	
\delta(y) = \int^{\infty}_{y} z P^2(z)dz, 
\end{eqnarray}
 which can easily be derived form the differential equation for $\delta$ and the condition $\delta(\infty)=0$. 
 
 Now we see that the coefficients before $ y^2 P^2(y)$, $ H^2(y)$ and $r_h^2 y^{2} V(\varphi)$ are all non-negative. Therefore, the left hand side of eq.(\ref{SCALINGCON}) can only vanish if 
 and only if $P^2(y)=H^2(y)=V(\varphi)=0$ which again proves the theorem. 

The above non-hair theorem was proven under a very weak condition on the potential, namely $V(\varphi)\ge 0$ and no other additional conditions on $V(\varphi)$ and the target space were imposed, which is rather desirable from a physical point of view.

\subsection{No-hair theorem for static, spherically symmetric black holes \\ in multiscalar theories with time dependent scalar fields} 

We focus again on static and spherically symmetric spacetimes with a metric given by (\ref{SSM}). This time we require the scalar fields to be spherically symmetric, $\partial_i\varphi^a=0$, but they can be time-dependent, i.e. can depend on both $r$ and $t$. The dependence on time has to be such that the energy-momentum tensor of the scalar fields to be time-independent. To guarantee that the energy-momentum tensor of the scalar fields is time independent we assume that the target space metric $\gamma_{ab}(\varphi)$ admits a Killing field $k^a$ whose flow leaves the potential invariant, namely ${\cal L}_{k}V(\varphi)=k^a\partial_a V(\varphi)=0$ and the dynamics of scalar fields is confined on the flow of $k^a$ \cite{Yazadjiev:2024rql}. To guarantee that the Jordan frame metric is also static we should assume that the conformal factor $A(\varphi)$ is invariant under the flow of $k^a$, i.e. ${\cal L}_{k}A(\varphi)=k^a\partial_a A(\varphi)=0$ 

From mathematical point of view we require the Lie derivative ${\cal L}_{\xi} \varphi^a$ along the timelike Killing field $\xi=\frac{\partial}{\partial t}$ satisfies 
${\cal L}_{\xi} \varphi^a = -\omega k^a$ where $\omega$ is a real number. In particular, if the orbits of the Killing field $k^a$ are periodic, then the time dependence will also be periodic in time. This is the case for example with boson stars which constitute of two scalar fields periodic in time or equivalently one periodic in time complex scalar field. 

From now on we assume the existence of such a Killing field $k^a$ and, in addition, we assume that the axis of $k^a$, i.e. the points where $k^a=0$, is non-empty. The norm of $k^a$ will be denoted by $|k|$, i.e. $|k|^2=\gamma_{ab}(\varphi)k^ak^b$.

With our assumptions in mind, the dimensionally reduced field equations are the following 
\begin{eqnarray}
	&&\frac{2}{r}e^{-2\Lambda} \Lambda^{\prime} + \frac{1}{r^2}\left(1-e^{-2\Lambda}\right)= \omega^2 e^{-2\Phi}|k|^2 + 
	e^{-2\Lambda}\gamma_{ab}(\varphi)\partial_r\varphi^a\partial_r\varphi^b + 2V(\varphi), \label{DRE} \\ \notag \\
	&&\frac{2}{r}e^{-2\Lambda} \Phi^{\prime} - \frac{1}{r^2}\left(1-e^{-2\Lambda}\right)= 
	\omega^2 e^{-2\Phi}|k|^2 + 
	e^{-2\Lambda}\gamma_{ab}(\varphi)\partial_r\varphi^a\partial_r\varphi^b - 2V(\varphi), \\
	&&e^{-2\Lambda}\left[\Phi^{\prime\prime} + (\Phi^{\prime} + \frac{1}{r})(\Phi^{\prime} - \Lambda^{\prime})\right]=
	-\frac{1}{2}K - 2V(\varphi), \\
	&&\partial_r \left(e^{\Phi-\Lambda} r^2 \gamma_{ab}(\varphi)\partial_r\varphi^b \right) = 
	\frac{1}{4}e^{\Phi + \Lambda} r^2 \left(\frac{\partial V(\varphi)}{\partial \varphi^a} + \frac{\partial K}{\partial \varphi^a} \right) 
 + \omega^2 r^2 e^{\Lambda -\Phi} k^b \frac{\partial k_a}{\partial \varphi^b},\label{EQFF} 
\end{eqnarray} 
where $K$ is defined by 
\begin{eqnarray}\label{K}
	K= 2\gamma_{ab}(\varphi)\nabla_{\mu}\varphi\nabla^{\mu}\varphi = -2\omega^2 e^{-2\Phi}|k|^2 + 
	2e^{-2\Lambda}\gamma_{ab}(\varphi)\partial_r\varphi^a\partial_r\varphi^b. 
\end{eqnarray} 

Taking into account that the metric functions $\Phi$ and $\Lambda$ depend on $r$ only and using the dimensionally reduced field equations it follows that $\gamma_{ab}(\varphi)\partial_r\varphi^a \partial_r\varphi^b $ and $|k|$ are functions of $r$ only. Then we can define
\begin{eqnarray}
	P^2(r)=\gamma_{ab}(\varphi)\partial_r\varphi^a \partial_r\varphi^b .
\end{eqnarray}

Using the asymptotic flatness and the dimensionally reduced field equations, one can easily show that 
\begin{eqnarray}
	\lim_{r\to \infty} P^2(r)=0, \;\; \; \lim_{r\to \infty} V(\varphi)=0, \;\;\; \lim_{r\to \infty} |k|=0, 
\end{eqnarray}
and more precisely all these quantities drop off at least as $\sim \frac{1}{r^4}$ for $r\to \infty$.
On the regular horizons the function $(\Phi + \Lambda)$ and its derivative are regular. Using this fact, and by adding the equation for $\Phi$ to the equation for $\Lambda$,
one can see that 
\begin{eqnarray}
	\lim_{r\to r_h} e^{-2\Phi}|k|^2=0,
\end{eqnarray}
and $e^{2\Lambda-2\Phi}|k|^2$ is finite on the horizon.

As in the previous section, the strategy to prove the no-hair theorem is based on a proper divergence identity. The desired divergence identity can be derived in the following way. We multiply the equation (\ref{EQFF}) for $\varphi^a$ by $\partial_r\varphi^a$ and after some algebra we obtain 
\begin{eqnarray}\label{I12}
	\partial_r \left(e^{\Phi-\Lambda} r^2 \gamma_{ad}(\varphi)\partial_r\varphi^a \partial_r\varphi^d \right) - 
	\left(e^{\Phi-\Lambda} r^2 \gamma_{ad}(\varphi) \partial_r\varphi^d \right) \partial^2_r\varphi^a \nonumber \\ = 
	\frac{1}{4}e^{\Phi + \Lambda} r^2 \left(\frac{\partial V(\varphi)}{\partial \varphi^a} + \frac{\partial K}{\partial \varphi^a} \right) \partial_r\varphi^a +\omega^2 r^2 e^{\Lambda -\Phi} k^b \frac{\partial k_a}{\partial \varphi^b} \partial_r\varphi^a . 
\end{eqnarray}
We then express $e^{\Phi-\Lambda} r^2 \partial_r^2 \varphi^a$
again from (\ref{EQFF}) and substitute into the above equation (\ref{I12}). In this way we get
\begin{eqnarray}\label{I22}
	\partial_r \left(e^{\Phi-\Lambda} r^2 \gamma_{ad}(\varphi)\partial_r\varphi^a \partial_r\varphi^d \right) + 
	\partial_r\left( r^2 e^{\Phi-\Lambda}\gamma_{ad}(\varphi)\right) \partial_r\varphi^a \partial_r\varphi^d \nonumber \\ = 
	\frac{1}{2}e^{\Phi + \Lambda} r^2 \left(\frac{\partial V(\varphi)}{\partial \varphi^a} + \frac{\partial K}{\partial \varphi^a} \right) \partial_r\varphi^a + 2\omega^2 r^2 e^{\Lambda -\Phi} k^b \frac{\partial k_a}{\partial \varphi^b} \partial_r\varphi^a . 
\end{eqnarray}
Further, making use of the explicit form (\ref{K}) of $K$ and after long algebra we find 
\begin{eqnarray}\label{I32}
	&&\frac{d}{dr}\left[ 2 r^2 e^{\Phi+\Lambda}V(\varphi) - r^2 e^{\Phi-\Lambda} P^2 - \omega^2 r^2 e^{\Lambda-\Phi} |k|^2 \right] \\
	&&=r e^{\Phi + \Lambda} \left[4V(\varphi) + (1+ e^{-2\Lambda})P^2 + \omega^2 e^{-2\Phi} |k|^2 (e^{2\Lambda} -3) \right].\nonumber 
\end{eqnarray}
The obtained divergence identity is however not what we want since the right-hand side is not manifestly non-negative. Another divergence identity with the desired property can be derived by multiplying the above identity by $r^{-2}$ and after some manipulation we get the desired result 
\begin{eqnarray}\label{I32}
	&&\frac{d}{dr}\left[ 2 e^{\Phi+\Lambda}V(\varphi) - e^{\Phi-\Lambda} P^2 - \omega^2 e^{\Lambda-\Phi} |k|^2 \right] \\
	&&= \frac{1}{r}e^{\Phi + \Lambda} \left[(1+ 3e^{-2\Lambda})P^2 + \omega^2 e^{2\Lambda -2\Phi} |k|^2 (1- e^{-2\Lambda}) \right].\nonumber 
\end{eqnarray}
The right-hand side of the above identity is non-negative which follows from the fact that $(1-e^{-2\Lambda})>0$ for $r\in [r_H,\infty)$ and this can be seen from 
the equation for $\Lambda$ written in the form 
\begin{eqnarray}
	\frac{d}{dr}\left[r(1-e^{-2\Lambda})\right] = r^2\left[\omega^2 e^{-2\Phi}|k|^2 + 
	e^{-2\Lambda}\gamma_{ab}(\varphi)\partial_r\varphi^a\partial_r\varphi^b + 2V(\varphi)\right]\ge 0.
\end{eqnarray}
Therefore $r(1-e^{-2\Lambda})$ is a non-decreasing function and taking into account that it is positive on the horizon and $r>0$, we conclude that $(1-e^{-2\Lambda})>0$ 
for $r\in[r_{H},+\infty)$.

The last step that remains to be made is to integrate identity (\ref{I32}) from the horizon to infinity and we get 
\begin{eqnarray}\label{I3}
	- 2e^{(\Phi + \Lambda)_{h}} V(\varphi_{h})	= \int^{+\infty}_{r_H} dr\frac{1}{r}e^{\Phi + \Lambda} \left[(1+ 3e^{-2\Lambda})P^2
 + \; \omega^2 e^{2\Lambda -2\Phi} |k|^2 (1- e^{-2\Lambda}) \right].\nonumber 
\end{eqnarray}
The left-hand side is non-positive by the assumption $V(\varphi)\ge 0$, while the right-hand side is non-negative. Therefore we conclude that $P(r)=0$ and $\omega^2|k|^2=(\partial_t\varphi)^2=0$ which means that the scalar fields are constant in the domain of outer communications. 

Summarizing, we proved the following theorem \cite{Yazadjiev:2024rql}
\medskip
\noindent

{\bf Theorem } {\it Let us consider the vacuum equations of the classical multiscalar theories of gravity and assume that the scalar field potential is non-negative, $V(\varphi)\ge 0$. Assume further that the scalar fields are invariant under the flow of the Killing fields generating the spherical symmetry, the target space metric admits a Killing field $k^a$ with a non-empty axis and the time dependence of the scalar fields satisfies ${\cal L}_{\xi}=-\omega k^a$ with $\omega$ being a real number. Then every static and spherically symmetric black hole solution to the field equations (\ref{FE1}) with regular horizon consists of the Schwarzschild solution and a constant map $\varphi_0$ with $V(\varphi_0)=k^2(\varphi_0)=0$. } 
\medskip
\noindent

\medskip
\noindent

This theorem, in the particular case of even-dimensional flat target space, reduces to the no-hair theorem for complex scalars given in \cite{Pena:1997cy}. In the case of target spaces with an odd number of dimensions, for example, the theorem excludes the existence of static, spherically black holes in the Friedberg - Lee - Sirlin model \cite{Friedberg:1976me} which contains one complex scalar field (depending harmonically on time) and one real scalar field.

We already know that the multiscalar theories with static scalar fields admit no solutions with a regular center. However, when the scalar fields are time-dependent, the multiscalar theories do admit such solutions which are, in fact, no-topological solitons (boson star solutions) \cite{Jetzer:1991jr, Liebling:2012fv, Collodel:2019uns}.

\subsection{Rotating black holes in multiscalar theories with scalar fields sharing spacetime symmetries and $V(\varphi)=0$ }

Here we consider stationary and axisymmetric black holes in vacuum multiscalar theories with $V(\varphi)=0$. We shall show that the only
stationary and axisymmetric, asymptotically flat black hole solution with a regular event horizon is the Kerr metric \cite{Heusler:1995qj}. 

Our assumptions are that both the spacetime and the scalar fields are stationary and axisymmetric. Concerning the scalar fields, this means that they are invariant under the flow of the stationary $\xi$ and axial $\eta$ Killing field,
\begin{eqnarray}
	{\cal L}_{\xi}\varphi^a={\cal L}_{\eta}\varphi^a=0.
\end{eqnarray}	
Under these assumptions, it can be shown that the spacetime is circular and the metric can be written in the form given in (\ref{RBHGR}). In addition, taking into account that 
\begin{eqnarray}
	R_{IJ}=R_{\mu\nu}K^{\mu}_I K^{\nu}_{J}=\gamma_{ab}(\varphi) (K^{\mu}_{I}\nabla_{\mu}\varphi^a) ( K^{\nu}_{J}\nabla_{\nu}\varphi^b) = 0,	
\end{eqnarray}
where $K_1=\xi$ and $K_2=\eta$, one can show that $\rho=\sqrt{-\det \Gamma_{IJ}}$ is harmonic on the factor space ${\hat {\cal M}}$ (see section (\ref{RBHGR})) 
\begin{eqnarray}
	{\hat D}_{i}{\hat D}^{i}\rho=\Gamma^{IJ}R_{IJ}=0. 
\end{eqnarray}	
The as we discussed in section (\ref{RBHGR}), $\rho$ and its conjugated harmonic function $z$ may be chosen as global coordinates on ${\hat {\cal M}}$ and the spacetime metric takes the standard Papapetrou form 
\begin{eqnarray}
	ds^2= -\frac{\rho^2}{X(\rho,z)} dt^2 + X(\rho,z)(d\phi + A(\rho,z)dt)^2 + \frac{e^{2h(\rho,z)}}{X(\rho,z)}(d\rho^2 + dz^2).
\end{eqnarray}	 
In the case under consideration, the dimensionally reduced equations of the multiscalar theories read 
\begin{eqnarray}\label{VEXCHI}
	&&\rho^{-1} \partial_{\rho}\left(\rho \partial_{\rho}X\right)
	+ \partial^2_{z} X = {1\over X} \left[\left(\partial_{\rho} X\right)^2 + \left(\partial_{z} X\right)^2 -
	4 \left(\partial_\rho\chi\right)^2 - 4 \left(\partial_z\chi\right)^2 \right], \notag	 \\ 
	&&\rho^{-1} \partial_{\rho}\left(\rho \partial_\rho\chi\right) +
	\partial^2_{z} \chi= {2\over X}\left[\partial_\rho\chi\partial_{\rho} X +
	\partial_z\chi \partial_{z} X \right] , 
\end{eqnarray}	
\begin{eqnarray}\label{VESF}	
	&&\rho^{-1} \partial_{\rho}\left(\rho \partial_{\rho}\varphi^a\right) + \gamma^{a}_{bc}(\varphi) [\partial_\rho\varphi^a \partial_\rho\varphi^b +\partial_z\varphi^a \partial_z\varphi^b ]=0, \;\;\; \;\;\;\;
\end{eqnarray}
together with the equations for $h$	
\begin{eqnarray}\label{}
	&&\rho^{-1}\partial_{\rho}h= {1\over 4X^2}\left[(\partial_{\rho}X)^2
	- (\partial_{z}X)^2 + 4 (\partial_\rho\chi)^2 - 4 (\partial_z\chi)^2\right] \notag \\ 
	&& + \gamma_{ab}(\varphi)(\partial_\rho\varphi^a\partial_\rho\varphi^b -\partial_z\varphi^a\partial_z\varphi^b) , \notag \\ \notag \\
	&&\rho^{-1}\partial_{z}h = {1\over 2X^2}\left[\partial_{\rho}X\partial_{z}X +
	4 \partial_\rho\chi \partial_z\chi\right] + \; 2\gamma_{ab}(\varphi)\partial_\rho\varphi^a\partial_z\varphi^b, \\ \notag \\ 
	&&\partial^2_\rho h + \partial^2_z h = - \frac{1}{4X^2} \left[(\partial_\rho X)^2 + (\partial_z X)^2 + 4 (\partial_\rho \chi)^2 + 4(\partial_z \chi)^2 \right]
	\notag \\ && - \gamma_{ab}(\varphi)(\partial_\rho\varphi^a\partial_\rho\varphi^b + \partial_z\varphi^a\partial_z\varphi^b)\nonumber 
\end{eqnarray}
and for the metric function $A$ 	 
\begin{eqnarray}\label{}
	&&\rho^{-1}\partial_{\rho}A = {2\over X^2} \partial_z\chi , \\ 
	&&\rho^{-1}\partial_{z}A = - {2\over X^2} \partial_\rho \chi. \nonumber
\end{eqnarray}
Let us recall that $\chi$ is the twist potential associated with the axial Killing vector $\eta$ and defined in section (\ref{RBHGR}). As clearly seem, the systems of equations (\ref{VEXCHI}) and (\ref{VESF}) are decoupled and may be solved independently. The key observation is that the equations for the metric function $h$ are 
linear which suggests the following partition 
\begin{eqnarray}
	h=h_{vac} + h_{\varphi},
\end{eqnarray}	
where $h_{vac}$ satisfies the equations
\begin{eqnarray}\label{}
	\rho^{-1}\partial_{\rho}h_{vac}&=& {1\over 4X^2}\left[(\partial_{\rho}X)^2
	- (\partial_{z}X)^2 + 4 (\partial_\rho\chi)^2 - 4 (\partial_z\chi)^2\right] \notag \\
	\rho^{-1}\partial_{z}h_{vac} &=& {1\over 2X^2}\left[\partial_{\rho}X\partial_{z}X +
	4 \partial_\rho\chi \partial_z\chi\right], \\ \notag \\
	\partial^2_\rho h_{vac} + \partial^2_z h_{vac} &=& - \frac{1}{4X^2} \left[(\partial_\rho X)^2 + (\partial_z X)^2 + 4 (\partial_\rho \chi)^2 + 4(\partial_z \chi)^2 \right],
	\nonumber 
\end{eqnarray}
while $h_{\varphi}$ satisfies 
\begin{eqnarray}\label{}
	\rho^{-1}\partial_{\rho}h_{\varphi}&=& \gamma_{ab}(\varphi)(\partial_\rho\varphi^a\partial_\rho\varphi^b -\partial_z\varphi^a\partial_z\varphi^b) , \notag \\ \notag \\
	\rho^{-1}\partial_{z}h_{\varphi} &=& 2\gamma_{ab}(\varphi)\partial_\rho\varphi^a\partial_z\varphi^b , \\ \notag \\
	\partial^2_\rho h_{\varphi} + \partial^2_z h_{\varphi} &=& - \gamma_{ab}(\varphi)(\partial_\rho\varphi^a\partial_\rho\varphi^b + \partial_z\varphi^a\partial_z\varphi^b).\nonumber 
\end{eqnarray} 

Let us note that the finiteness of the Ricci scalar and the regularity of the derivatives of $\varphi$ guarantee that $h_{\varphi}$ and its derivatives are finite on the boundary of ${\hat {\cal M}}$. Concerning the behavior of $h_\varphi$ near infinity, the asymptotic flatness implies 
\begin{eqnarray}\label{ABH}
	h_{\varphi}=O(1/r^2)
\end{eqnarray}
for $r\to \infty$ where $r=\sqrt{r^2 + z^2}$.

The last crucial step is to consider the following vector field defined on the factor space ${\hat {\cal M}}$:
\begin{eqnarray}
	v_i=\rho {\hat D}_i e^{-h_{\varphi}}. 
\end{eqnarray}
Applying the Gauss theorem to the divergence of this vector field on ${\hat {\cal M}}$, we get 
\begin{eqnarray}\label{DISMSTT}
	&&\int_{\partial{\hat M}\cup \infty} \rho e^{-h_{\varphi}} \left[\partial_z h_{\varphi} d\rho - \partial_{\rho}h_{\varphi}dz \right] 	\\ 
	&&= \int_{{\hat M}} \rho e^{-h_{\varphi}} \left[(\partial_{\rho}h_{\varphi})^2 + (\partial_{\rho}h_{\varphi})^2 - \left[\frac{1}{\rho}\partial_{\rho}h_{\varphi} + \partial^2_\rho h_{\varphi} + \partial^2_z h_{\varphi}\right]\right] d\rho dz\notag,
\end{eqnarray}
where $\infty$ in the left-hand side integral means integration over the semi-circle at infinity. Using the equations for $h_{\varphi}$ we easily find that
that 
\begin{eqnarray}
	\frac{1}{\rho}\partial_{\rho}h_{\varphi} + \partial^2_\rho h_{\varphi} + \partial^2_z h_{\varphi}=-2\gamma_{ab}(\varphi)\partial_z\varphi^a\partial_{z}\varphi^b, 
\end{eqnarray}	 
and then the integral identity above takes the form 
\begin{eqnarray}\label{DISMSTT1}
	&&\int_{\partial{\hat M}\cup \infty} \rho e^{-h_{\varphi}} \left[\partial_z h_{\varphi} d\rho - \partial_{\rho}h_{\varphi}dz \right] 	\\ 
	&&= \int_{{\hat M}} \rho e^{-h_{\varphi}} \left[(\partial_{\rho}h_{\varphi})^2 + (\partial_{\rho}h_{\varphi})^2 + 2\gamma_{ab}(\varphi)\partial_z\varphi^a\partial_{z}\varphi^b\right] d\rho dz\notag.
\end{eqnarray} 
The right-hand side of the integral identity is now manifestly non-negative. Our last task is to show that the boundary integral on $\partial{\hat M}\cup \infty$ vanishes. $h_{\varphi}$ and its derivatives in $\rho$ and $z$ are finite on $\partial{\hat M}$ which, combined with the fact that $\rho$ vanishes on $\partial{\hat M}$, gives 
\begin{eqnarray} 
	\int_{\partial{\hat M}} \rho e^{-h_{\varphi}} \left[\partial_z h_{\varphi} d\rho - \partial_{\rho}h_{\varphi}dz \right]=0. 
\end{eqnarray}

It remains to estimate the boundary integral over the semi-circle at infinity. In the asymptotic region, it is convenient to introduce the coordinates $r$ and $\theta$ defined by $\rho=r\sin\theta$ and $z=r\cos\theta$. In therm of these coordinates, we have 
\begin{eqnarray} 
	\int_{\infty}\! \!\rho e^{-h_{\varphi}} \!\left[\partial_z h_{\varphi} d\rho - \partial_{\rho}h_{\varphi}dz \right] = - \lim_{r\to \infty}\int^{\pi}_{0} r^2 (e^{-h_{\varphi}}\partial_{r} h_{\varphi}) \sin\theta d\theta = 0, \notag
\end{eqnarray}
where we have taken into account (\ref{ABH}). Thus, we showed that the left hand side of (\ref{DISMSTT1}) is zero, implying that both non-negative integrands on the right-hand side vanish. Hence, $h_{\varphi}$ is constant in the entire domain of outer communications and taking into account that $\lim_{r\to \infty}h=0$ we conclude that $h_{\varphi}=0$. This means that the scalar fields are constant in the domain of outer communications. In this way, we proved the following 

\medskip
\noindent

{\bf Theorem} {\it Let us consider the vacuum equations of the classical multiscalar theories of gravity with $V(\varphi)=0$. Then every stationary and axisymmetric black hole solution to the field equations with stationary and axisymmetric scalar fields, ${\cal L}_{\xi}\varphi^a={\cal L}_{\eta}\varphi^a=0$, and regular horizon consists of the Kerr solution and a constant multiscalar map. }

\medskip
\noindent

\subsection{No-hair theorem for rotating black holes with scalar fields sharing the spacetime symmetries and $V(\varphi)\ne 0$}

In this section, we consider stationary and axisymmetric black holes in vacuum multiscalar theories with $V(\varphi)\ne 0$. We assume that both the spacetime and the scalar fields are stationary and axisymmetric, i.e. for the scalar fields we impose 
\begin{eqnarray}
	{\cal L}_{\xi}\varphi^a={\cal L}_{\eta}\varphi^a=0.
\end{eqnarray}	
We will present a new theorem that, to the best of our knowledge, has not been proven in the literature so far. 

The covariant derivative with respect to the target space metric $\gamma_{ab}(\varphi)$ on ${\cal E}_{N}$ we denote by ${\cal D}_a$. Our starting point is the scalar fields equation 
\begin{eqnarray}
	\nabla_{\mu}\nabla^{\mu}\varphi^a + \gamma^{a}_{bc}(\varphi)\nabla_{\mu}\varphi^b\nabla^{\mu}\varphi^c =\frac{1}{4}\gamma^{ab}(\varphi)\frac{\partial V(\varphi)}{\partial \varphi^b}, 
\end{eqnarray}	
which we multiply by $ {\cal D}_{a}V(\varphi)=\frac{\partial V(\varphi)}{\partial \varphi^a}$ and obtain 
\begin{eqnarray}
	{\cal D}_{a}V(\varphi)\nabla_{\mu}\nabla^{\mu}\varphi^a + {\cal D}_{a}V(\varphi)\gamma^{a}_{bc}(\varphi)\nabla_{\mu}\varphi^b\nabla^{\mu}\varphi^c =\frac{1}{4}\gamma^{ab}(\varphi){\cal D}_{a}V(\varphi){\cal D}_{b}V(\varphi). 
\end{eqnarray}
Taking into account that 
\begin{eqnarray}
	{\cal D}_{a}V(\varphi)\nabla_{\mu}\nabla^{\mu}\varphi^a = \nabla_{\mu} ({\cal D}_{a}V(\varphi)\nabla^{\mu}\varphi^a)- \frac{\partial^2 V(\varphi)}{\partial \varphi^b \partial \varphi^c} \nabla_{\mu}\varphi^b \nabla^{\mu}\varphi^c 
\end{eqnarray}
we find that 
\begin{eqnarray}
	&&\nabla_{\mu} ({\cal D}_{a}V(\varphi)\nabla^{\mu}\varphi^a)- \left[\frac{\partial^2 V(\varphi)}{\partial \varphi^b \partial \varphi^c} - \gamma^{a}_{bc}(\varphi) {\cal D}_a V(\varphi)\right] \nabla_{\mu}\varphi^b \nabla^{\mu}\varphi^c \notag \\ 
	&&=\frac{1}{4}\gamma^{ab}(\varphi){\cal D}_{a}V(\varphi){\cal D}_{b}V(\varphi).
\end{eqnarray}
The term in the squared brackets is just ${\cal D}_b{\cal D}_c V(\varphi)$ and therefore we finally get
\begin{eqnarray}\label{MSDIRBH}
	\nabla_{\mu} ({\cal D}_{a}V(\varphi)\nabla^{\mu}\varphi^a)
	= {\cal D}_b{\cal D}_c V(\varphi) \nabla_{\mu}\varphi^b \nabla^{\mu}\varphi^c + \frac{1}{4}\gamma^{ab}(\varphi){\cal D}_{a}V(\varphi){\cal D}_{b}V(\varphi).
\end{eqnarray}
With this divergence identity at hand, we are ready to prove the following 

\medskip
\noindent

{\bf Theorem} {\it Let us consider the vacuum equations with of the classical multiscalar theories of gravity with $V(\varphi)\ne 0$. Assume that the potential $V(\varphi)$ is such that the tensor ${\cal D}_b{\cal D}_c V(\varphi)$ is semi-positive definite everywhere on ${\cal E}_N$. Then every stationary and axisymmetric black hole solution to the field equations with stationary and axisymmetric scalar fields, ${\cal L}_{\xi}\varphi^a={\cal L}_{\eta}\varphi^a=0$, and regular horizon consists of the Kerr solution and a constant scalar map with ${\cal D}_aV(\varphi)=0$. }

\medskip

We shall prove the theorem by reducing the problem to a pure Riemannian problem. The first step is to notice the spacetime is circular. Indeed, it is not difficult to check that circularity conditions 
$\xi^\mu T_{\mu[\nu}\xi_\alpha\eta_\beta]=0$ and $\eta^\mu T_{\mu[\nu}\eta_\alpha\xi_\beta]=0$ are satisfied as a consequence of the fact that ${\cal L}_{\xi}\varphi^a={\cal L}_{\eta}\varphi^a=0$. 
Next, due to the spacetime isometries the identity (\ref{MSDIRBH}) is equivalent to the following elliptic identity on the factor space $({\hat {\cal M}}, {\hat g}_{ij})$ (see sec.(\ref{RBHGR})) 
\begin{eqnarray}\label{RED_MSDIRBH}
	{\hat D}_i (\rho{\cal D}_{a}V(\varphi) {\hat D}^i \varphi^a)
	= \rho\left[{\cal D}_b{\cal D}_c V(\varphi) {\hat D}_i\varphi^b {\hat D}^i\varphi^c + \frac{1}{4}\gamma^{ab}(\varphi){\cal D}_{a}V(\varphi){\cal D}_{b}V(\varphi)\right].
\end{eqnarray}
Here $\rho$ is just the square root of the absolute value of the determinant of the Gram matrix of the Killing fields given in eq.(\ref{DEF_rho}). Integrating identity (\ref{RED_MSDIRBH}) over 
${\hat {\cal M}}$ and applying the Gauss theorem for the left-hand side we obtain 
\begin{eqnarray}\label{INT_RED_MSDIRBH}
&&	\int_{\partial {\hat {\cal M}} \cup \infty} \rho{\cal D}_{a}V(\varphi) {\hat D}_i \varphi^a dS^{i}
\\ &&	= \int_{{\hat {\cal M}}}\rho\left[{\cal D}_b{\cal D}_c V(\varphi) {\hat D}_i\varphi^b {\hat D}^i\varphi^c + \frac{1}{4}\gamma^{ab}(\varphi){\cal D}_{a}V(\varphi){\cal D}_{b}V(\varphi)\right] \sqrt{{\hat g}}d^2x, \notag 
\end{eqnarray}
where the integral over the boundary includes also an integration over the semi-circle at infinity. The boundary integral over $\partial {\hat {\cal M}}$ vanishes 
since $\partial {\hat {\cal M}}$ consists of the finite interval $I_H$ corresponding to the horizon and the two semi-infinite intervals $I_{+}$ and $I_{-}$ and corresponding to the axis of $\eta$ where $\rho$ vanishes as we discussed in sec.(\ref{RBHGR}). The boundary integral over the infinite semi-circle also vanishes due to the fact at infinity $V(\varphi)$ drops off rapidly enough as dictated by 
the asymptotic flatness. Therefore, we conclude that the right-hand side of (\ref{INT_RED_MSDIRBH}) vanishes, too. Under conditions we imposed on the potential $V(\varphi)$, this is possible only when 
${\hat D}_i\varphi^a=0$ and ${\cal D}_a V(\varphi)=0$, which proves the theorem.

\section{Black holes with a noncanonical scalar field}

Here we focus on a scalar-tensor theory with non-canonical scalar fields given by the action (\ref{HordeskiNCSF}). 
The scalar field equation of motion yielded by this action is 
\begin{eqnarray}\label{NCSFE}
	\nabla_{\mu}(\partial_{K}F\nabla^{\mu}\varphi) + \partial_{\varphi}F=0 . 
\end{eqnarray}

The following black hole no-hair theorem holds \cite{Graham:2014mda} 

\medskip
\noindent

{\bf Theorem } { \it Consider an asymptotically flat, four-dimensional black hole spacetime which is either static or stationary and axisymmetric, with a minimally
	coupled scalar field having an action of the form (\ref{HordeskiNCSF}). We also assume that the scalar field shares the
	same symmetries as the spacetime metric. Then the only black hole solutions of the Einstein equations with this matter source are the Kerr or the Schwarzschild solution with $\varphi=constant$, provided that
	either} 
\begin{eqnarray}
	\partial_{K}F > 0 \;\; \; \text{and} \;\; \;\varphi\partial_{\varphi}F\ge 0
\end{eqnarray}

{\it or} 
\begin{eqnarray}
	\partial_{K}F < 0 \;\;\; \text{and} \;\;\; \varphi\partial_{\varphi}F\le 0.
\end{eqnarray}

\medskip
\noindent

As in the case of a canonical scalar field discussed in section \ref{VBHSSF}, we consider the region ${\cal D}$ bounded by the partial Cauchy surface ${\cal J}$, an isometric copy 
${\cal J}^{\prime}\ne {\cal J}$ of ${\cal J}$, a portion of the event horizon $H$ and a timelike 3-surface ${\cal T}$ at infinity. Next, we multiply Eq.(\ref{NCSFE}) by $\varphi$ and integrate over ${\cal D}$. After integrating the first term by parts we get the following identity 
\begin{eqnarray}
	\int_{{\cal D}} d^4x\sqrt{-g} \left(\partial_{K}F \nabla_{\mu}\varphi \nabla^{\mu}\varphi - \varphi \partial_{\varphi}F\right)=
	\int_{\partial {\cal D}} d^3S n^{\mu}\nabla_{\mu}\varphi (\varphi \partial_{K}F). \notag\\
\end{eqnarray}
The boundary integrals over ${\cal J}$ and ${\cal J}^{\prime}$ cancel one another just as for the canonical case. Since the spacetime is asymptotically flat, the contribution from infinity to the boundary term is also vanishing. The integral over the horizon vanishes, too. This follows from the fact that the scalar field shares the spacetime symmetries and the fact that the scalar field and
its first derivative are regular at the horizon (otherwise the components of the energy-momentum tensor would
be divergent at the horizon). As a consequence, our integral identity reduces to 
\begin{eqnarray}\label{NCSFID1}
	\int_{{\cal D}} d^4x\sqrt{-g} \left(\partial_{K}F \nabla_{\mu}\varphi \nabla^{\mu}\varphi - \varphi \partial_{\varphi}F\right)=0
\end{eqnarray}
Since the scalar field shares the spacetime symmetries, the gradients of the scalar field are spacelike or zero.
Thus, if $\partial_{K}F$ is of definite sign and $\varphi\partial_{\varphi}F$ is of
definite, but opposite sign, then Eq. (\ref{NCSFID1}) cannot be satisfied, and hence we conclude that no hairy black hole
solutions can exist but the trivial $\varphi=constant$. Indeed, when there is a potential term in the action the only solution will in general be $\varphi=0$. This completes the proof. In particular, if the action is purely kinetic with $\partial_{K}F$ of a fixed sign no such solutions can exist. It should be
noted that the proof also assumes implicitly that $\varphi\to 0$ as $r\to \infty$, so it does not cover Higgs-like potentials.

The above result can be extended to the case when the scalar field may have time dependence in the case of circular spacetimes \cite{Graham:2014ina}. It is clear that stationary black hole solutions with time-dependent scalar field could in general exist if the Lagrangian has no explicit dependence on $\varphi$, i.e. $F=F(K)$, otherwise, the scalar field energy-momentum tensor would depend on time. Using the field equations, it is easy to show that the scalar field can only depend on the coordinates $x^{I}=(t,\phi)$ associated with the Killing vectors. For the energy-momentum tensor to be still independent of time, the terms $F(K)$ and $\partial_K F (\partial_t\varphi)^2 $ have to be time independent. In the general case, this requires $\varphi$ to be linear in $t$. We can also rule out $\varphi$ depending upon $\phi$. This is because the spacetime is axisymmetric, so the energy-momentum tensor can only be consistent with the axisymmetry of the spacetime if $\varphi$ depends at most linearly upon $\phi$. However, since $\phi$ is a periodic coordinate clearly this cannot be the case, as otherwise, $\varphi$ would not be a continuous, single-valued function. Therefore, the scalar field can in fact only depend upon time and $\varphi=\alpha t + \beta $ with $\alpha$ and $\beta$ being constants. The final step is to show that only $\alpha=0$ is consistent with the asymptotically flatness. For this purpose, we consider the asymptotic behavior of the energy-momentum tensor. For $r\to \infty $ we have $K \to 2\alpha^2$ and 
\begin{eqnarray}
T_{tt}\to 2\alpha^2 \partial_KF(2\alpha^2) -\frac{1}{2} F(2\alpha^2),\; \; \;\;\; T_{rr}\to \frac{1}{2} F(2\alpha^2). 
\end{eqnarray}
Therefore we must have $\partial_KF(2\alpha^2)=0$ and $F(2\alpha^2)=0$, which, in the general case and for non-pathological actions, is possible only when $\alpha=0$.

\section{Black holes in shift-symmetric Horndeski theories}\label{}

Horndeski theories are rather complicated and a complete black hole solution classification within them is possible only under certain and strong enough restrictions. That is why we focus on Horndeski theories with shift symmetry \cite{Hui:2012qt, Sotiriou:2013qea}, i.e. theories invariant
under the transformation $\varphi\to \varphi + constant$. The most general theory of this kind can be derived by simply dropping the $\varphi$ dependence for $F$ and $G_i$
in the action (\ref{Hordeski_Action}).

For shift symmetric theories, the equation for the scalar field can be written as a current conservation
\begin{eqnarray}\label{CCHT}
	\nabla_{\mu}J^{\mu}=0,
\end{eqnarray}
with 
\begin{eqnarray}
	J^{\mu}= \frac{\partial {\cal L}}{\partial (\partial_{\mu}\varphi)}.
\end{eqnarray}
Let us focus on static and spherically symmetric spacetimes having a metric ansatz of the form
\begin{eqnarray}
	ds^2= - f(r)dt^2 + S(r)dr^2 + r^2(d\theta^2 + \sin^2\theta d\phi^2).
\end{eqnarray}
In addition, we shall assume that the scalar field shares the spacetime symmetries, i.e. it depends on the radial coordinate $r$ only.

As a first step, we shall show that the only non-zero component of the conserved current $J^\mu$ is $J^r$. Consider the Killing fields $\zeta_I$, $I=1,2,3$,
generating the spherical symmetry. Then, in accordance with our assumptions, we have 
\begin{eqnarray}
	{\cal L}_{\zeta_I} g_{\mu\nu}=0, \;\;\; {\cal L}_{\zeta_I} \varphi=0.
\end{eqnarray}
Any tensor constructed solely from the metric, the scalar field, and their derivatives, share the same symmetries. In particular, we have 
\begin{eqnarray}
	{\cal L}_{\zeta_I} J^{\mu}=0,
\end{eqnarray}
i.e. $J^\mu$ is invariant under the group of rotations $SO(3)$. Assuming that $J^{\mu}$ is not orthogonal to the 2-dimensional spheres $r=constant$, there will be a nonzero projection onto them. However, a non-zero vector field on the sphere cannot be invariant under all rotations, which contradicts the invariance of $J^\mu$ under the action of the $SO(3)$ group of isometries. Therefore, $J^\mu$ is orthogonal to the 2-dimensional spheres $r=constant$ and the angular components of $J^\mu$ are zero. Concerning the component $J^t$, it also has to vanish -- otherwise, it would select a preferred time direction. We are thus left with a radial component $J^r$ only. 

Further, from the conservation equation (\ref{CCHT}) we get 
\begin{eqnarray}
	\sqrt{f(r)S(r)}r^2J^r=constant. 
\end{eqnarray}
The next step is to show that the constant is in fact zero. We assume that $J^\mu J_\mu$ is regular on the horizon. Then, it is seen that, for $J^{\mu}J_{\mu}=(J^r)^2 S(r)$ to be regular there, $J^r$ has to vanish on the horizon. We then conclude that $J^r=0$ everywhere.

The last and trickiest step is to show that $J^r=0$ implies $\varphi=constant$. This relies on the actual dependence of the current on $\varphi$ and its derivatives. To perform the last step, we shall focus on theories for which the functions $G_i$ or their derivatives with respect to $K$ have no poles at $K\to 0$ \cite{Sotiriou:2013qea}. We also need the explicit form of $J^r$, which can be easily obtained, namely 
\begin{eqnarray}
	&&J^r = -2 \frac{\varphi^\prime}{S} \partial_{K}F + 2 \frac{rf^\prime + 4f}{rfS^2} \varphi^{\prime 2} \partial_{K} G_3 + 4\frac{f- fS + rf^\prime}{r^2fS^2}\varphi^\prime \partial_{K}G_4 \notag \\ 
	&& - 16 \frac{f+r f^\prime}{r^2fS^3} \varphi^{\prime 3} \partial^2_{K} G_{4} + 2 \frac{Sf^\prime - 3f^{\prime}}{r^2 f S^3} \varphi^{\prime 2} \partial_{K} G_{5}
	+ 8 \frac{f^\prime}{r^2fS^4} \varphi^{\prime 4} \partial^2_{K} G_{5}. \notag \\
\end{eqnarray}

From the above expression for $J^r$, we see that every term does depend at least linearly on $\varphi^\prime$. Additionally, assuming that $F$
has a piece linear in $K$ so that in the weak field limit the standard canonical kinetic term is present in the action,
the current is asymptotically proportional to $\varphi^\prime$, i.e. $\varphi^\prime$ vanishes asymptotically. The derivative $\varphi^\prime$ must vanish everywhere otherwise $J^r$ should be different from zero which is a contradiction. 

In this way, we have shown that the static, spherically symmetric black holes in the shift-symmetric Horndeski theories cannot sustain non-trivial scalar fields
under the assumptions we made \cite{Hui:2012qt, Sotiriou:2013qea}. 

Theories with functions $G_i$ having a pole at $K\to 0$ are in general more or less pathological with one notable exception. This is the theory with $F=K$, $G_3=0$, $G_4=1$ and $G_5=-4\alpha \ln(|K|)$ where $\alpha$ is a constant. This theory is in fact equivalent to the following particular case of the scalar-Gauss-Bonnet gravity
\begin{eqnarray}\label{SIFTSSGBG}
	S= \frac{1}{16\pi}\int d^4x\sqrt{-g} \left[R - 2\nabla_{\mu}\varphi\nabla^{\mu}\varphi + 4\alpha \varphi {\cal R}^2_{\;\;GB} \right].
\end{eqnarray}
The theory defined by the above action does admit black holes with a nontrivial scalar field \cite{Sotiriou:2013qea}. 

Scalar field endowed black holes can also be found in some particular shift-symmetric cases of the Horndeski theories provided the scalar field is allowed to be time-dependent \cite{Babichev:2013cya}. An explicit example is the following theory 
\begin{eqnarray}
	S= \frac{1}{16\pi G} \int d^4x\sqrt{-g}\left[R - \gamma \nabla_{\mu}\varphi \nabla^{\mu}\varphi + \beta G^{\mu\nu}\nabla_{\mu}\varphi\nabla_{\nu}\varphi -2\Lambda\right], 
\end{eqnarray}
where $G_{\mu\nu}$ is the Einstein tensor, $\gamma$ and $\beta$ are constants and $\Lambda$ is the cosmological constant. For this theory the conserved current is 	
\begin{eqnarray}
	J^{\mu}= (\gamma g^{\mu\nu} - \beta G^{\mu\nu})\nabla_{\nu}\varphi. 
\end{eqnarray}
It can be seen that the current can be eliminated by imposing $\gamma g_{\mu\nu}=\beta G_{\mu\nu}$ instead of $\varphi=constant$. Regular black hole solutions can then be found by allowing the scalar field to have a linear time dependence \cite{Babichev:2013cya}.
This is consistent with requiring the black hole spacetime to be spherical and static since the scalar field only enters the field equations through
its derivatives. In particular, for $\gamma=0 $ and $\Lambda=0$ a solution with precisely the Schwarzschild geometry and a nontrivial scalar field can be constructed, namely 
\begin{eqnarray}
	\varphi= qt\pm 2qM\left[2\sqrt{\frac{r}{2M}} + \ln \frac{\sqrt{r}-\sqrt{2M}}{\sqrt{r}+\sqrt{2M}}\right] + \varphi_0,
\end{eqnarray}	
where $q$ and $\varphi_0$ are constants. Depending on the chosen sign in the expression above, the scalar field will diverge at either the future or the past event horizon. Still, the geometry is regular on and outside the horizon. This construction was dubbed dressing a BH with a time-dependent Galileon.

Partial generalization of no-hair result for the case of rotating black holes can also be achieved \cite{Capuano:2023yyh}. We will present the result of \cite{Capuano:2023yyh} in a slightly modified form and we shall give a more concise and simpler proof. Consider shift-symmetric Horndeski theories for which the functions $F$, $G_i(K)$, and their derivatives are regular. We assume that the spacetime is circular and asymptotically flat and the scalar field shares the spacetime symmetries. As an additional assumption, we require that the Noether current $J_\mu$ in the asymptotic region has the form $J_{\mu} = \partial_{\mu}\varphi$. Under these assumptions, we shall show that black holes cannot have a scalar charge\footnote{The scalar charge $D$ is defined by the asymptotic behavior of the scalar field $\varphi=\frac{D}{r} + O(\frac{1}{r^2})$ for $r\to \infty$.}. In other words, the asymptotically flat, axisymmetric, and stationary black holes (with a circular spacetime) in shift-symmetric theories cannot develop a term $\frac{1}{r}$ in the scalar profile at large distances.
 
It is not difficult to see that under our assumptions we have $J_\mu \xi^{\mu}=J_\mu \eta^{\mu}=0$ where $\xi$ and $\eta$ are the stationary and the axial Killing field, respectively. Equivalently, $J_{\mu}$ can be viewed as a (co-)vector defied on the factor space $({\hat {\cal M}}, {\hat g}_{ij})$ (see sec.(\ref{RBHGR})). With this fact taken into account, one can easily show that eq. (\ref{CCHT}) reduces to the following equation on $({\hat {\cal M}}, {\hat g}_{ij})$ 
\begin{eqnarray}
	{\hat D}_{i}\left( \rho J^i\right)=0.
\end{eqnarray}	
Now, integrating on ${\hat {\cal M}}$ and applying the Gauss theorem we find 
\begin{eqnarray}
	\int_{\partial{\hat {\cal M}}\cup \infty}\rho J_i dS^i=\int_{\partial {\hat {\cal M}}}\rho J_i dS^i + \int_{\infty}\rho J_i dS^i =0,
\end{eqnarray}	
 where the boundary integral includes also integration over the semicircle at infinity. The boundary integral over $ \partial{\hat {\cal M}} $ vanishes because
 $ \partial{\hat {\cal M}}$ consists of the axis of the Killing field $\eta$ and the horizon where $\rho=0$. The last step is to explicitly calculate the integral over 
 the semi-circle at infinity. This can be done by introducing the standard asymptotic coordinates $(r, \theta)$ with $\rho=r\sin\theta$.Using also the fact that asymptotically $J_i=\partial_i\varphi$, we obtain 
 \begin{eqnarray}
 	0=\int_{\infty}\rho J_i dS^i= -\lim_{r\to \infty}\int^\pi_{0} r^2 \partial_r \varphi \sin\theta\, d\theta = 
 	2 \lim_{r\to \infty} r^2 \partial_r\varphi = -2 D,
 \end{eqnarray}	
 which means that the scalar charge $D$ is zero.

\section{Challenges and open problems}
There exist no general theorems classifying black holes with non-trivial scalar (or multiscalar) hair. As we mentioned in the introduction, the natural reason for the lack of such theorems is the complex mathematical nature of extended scalar-tensor gravity. 

To get some insight into the difficulties encountered in the attempts to classify the hairy black holes, we shall discuss some recently discovered solutions. More precisely we shall first focus on the scalar-Gauss-Bonnet gravity with action and field equations given by (\ref{ASGBG}) and \eqref{FESGBG_1}--\eqref{FESGBG_2}, respectively. The spectrum of the hairy black hole solutions within this theory depends strongly on the Gauss-Bonnet coupling function $f(\varphi)$. Of special interest are coupling functions that allow for the so-called spontaneous scalarization \cite{Doneva:2022ewd}. Such subclasses of scalar-Gauss-Bonnet theory are probably the best examples where one can demonstrate how involved and complex the classification problem can be and what kind of difficulties can be encountered. There is no need even to consider rotating solutions, the simplest static and spherically symmetric case reveals the complexity of the classification problem.

\begin{figure}[htb]
	\includegraphics[width=0.495\textwidth]{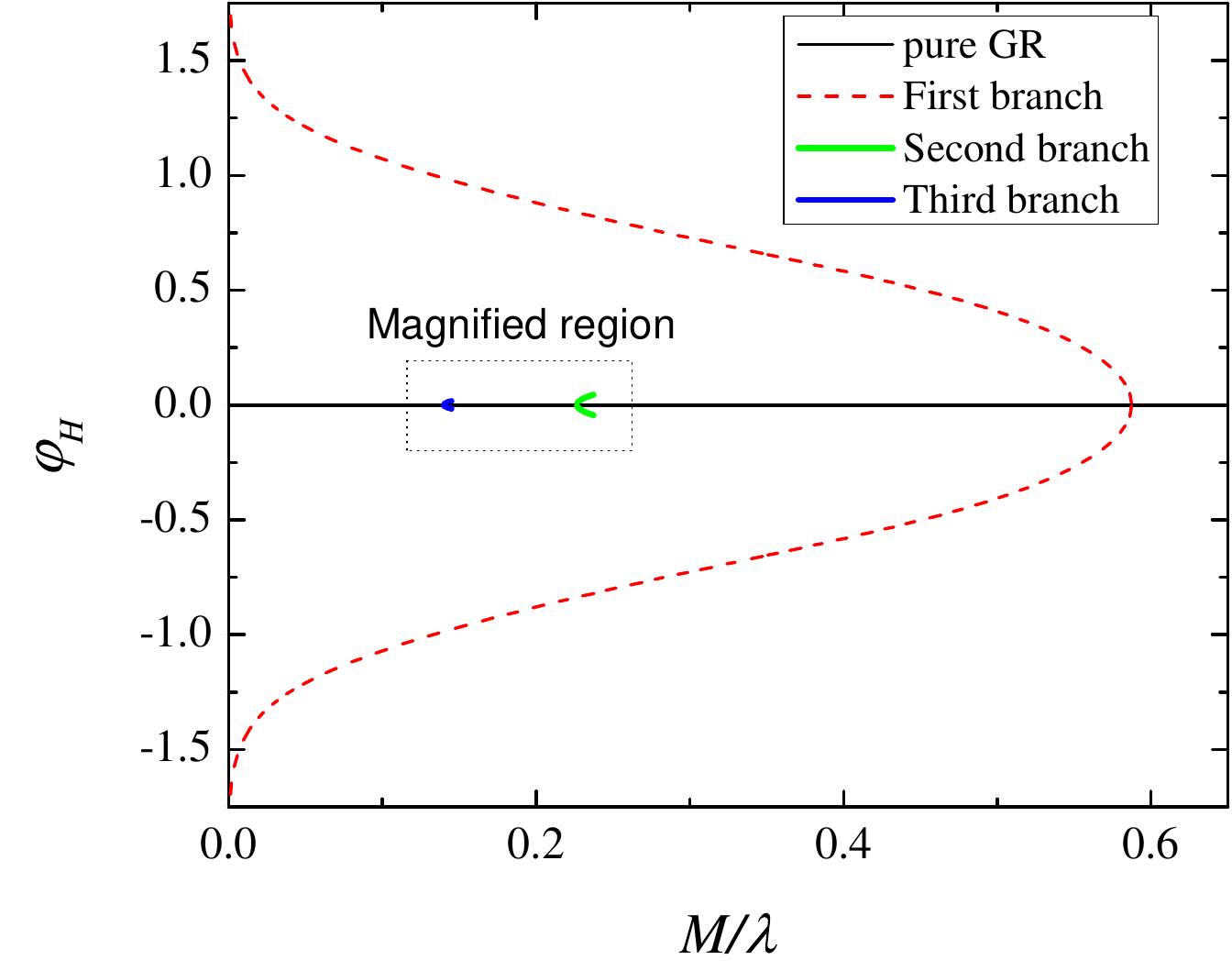}
	\includegraphics[width=0.495\textwidth]{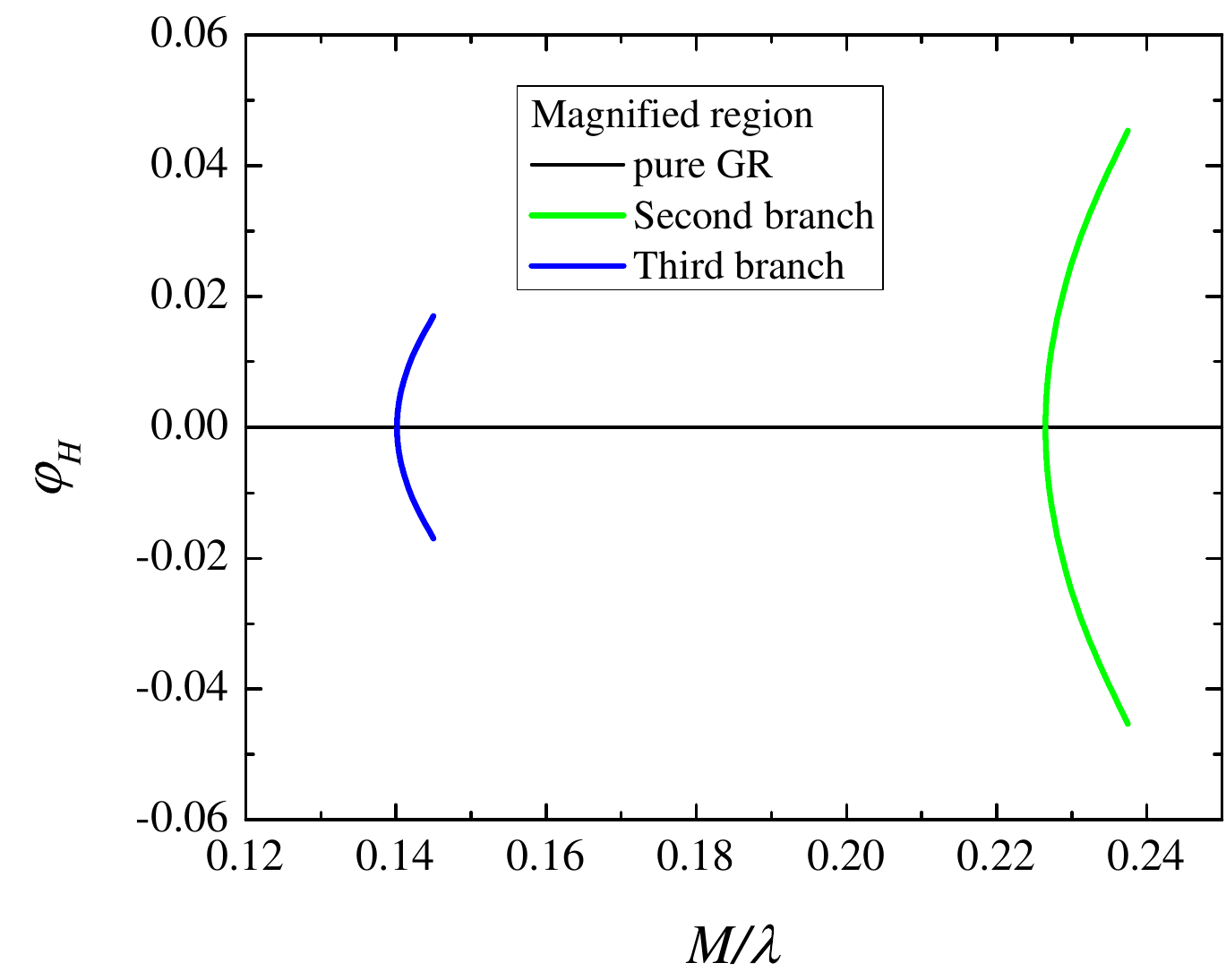}
	\caption{The scalar field at the horizon as a function of the normalized mass $M/\lambda$ for black holes within scalar-Gauss-Bonnet gravity with a coupling function \eqref{eq:coupling_function} having $\beta=6$. The right figure is a magnification of the left one. All nontrivial branches start from a bifurcation point at the trivial branch and they span either to $M=0$ (the first nontrivial branch) or they are terminated at some nonzero $M$ (all the other nontrivial branches)}
	\label{fig:phiH(M)}
\end{figure}

\begin{figure}[htb]
	\includegraphics[width=0.495\textwidth]{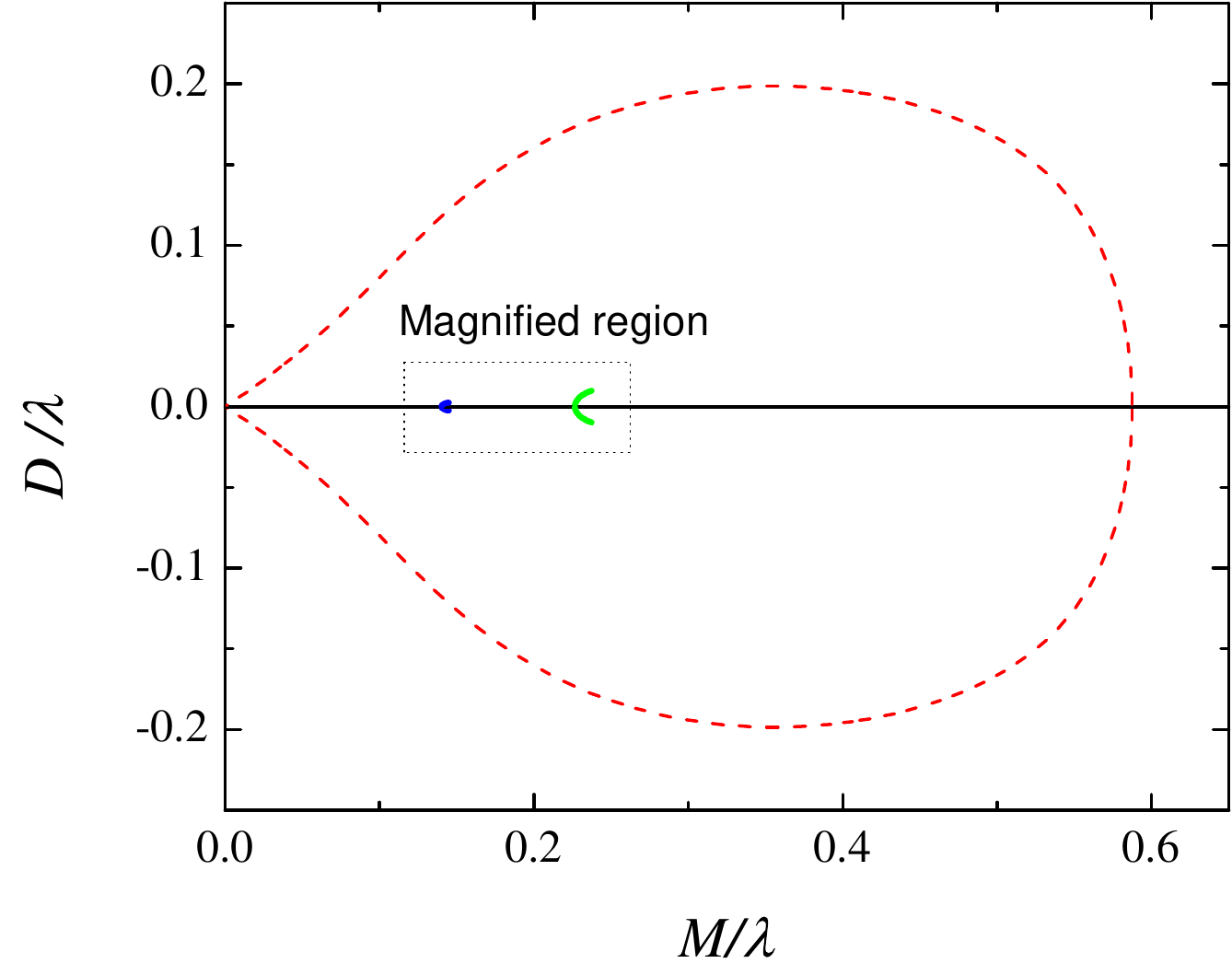}
	\includegraphics[width=0.495\textwidth]{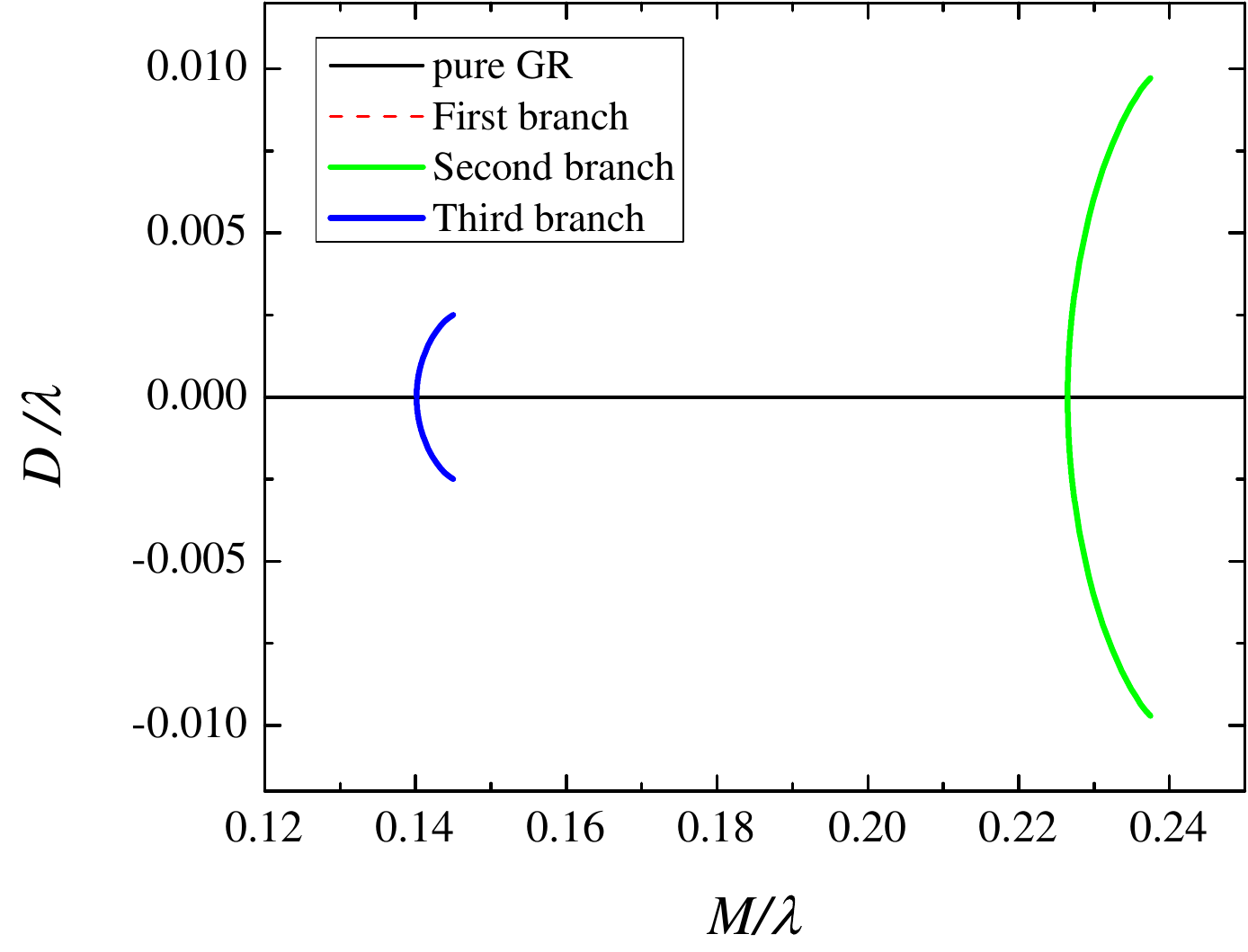}
	\caption{The scalar charge as a function of the normalized mass $M/\lambda$ for the same black hole solutions as in Fig. \ref{fig:phiH(M)}. The right figure is a magnification of the left one.}
	\label{fig:D(M)}
\end{figure}

To be specific, we consider the following coupling function 
\begin{equation}
	f(\varphi)= \frac{1}{12} \left[1- \exp(-6\varphi^2)\right], \label{eq:coupling_function}
\end{equation} 
which allows for spontaneous scalarization. Since $\frac{df}{d\varphi}(0)=0$, the Schwarzschild black hole is always a solution of the scalar-Gauss-Bonnet field equations with a zero scalar field $\varphi=0$. However, below a certain value of the black hole mass, the Schwarzschild solution becomes linearly unstable in the framework of the scalar-Gauss-Bonnet theory under consideration and new solutions with a non-trivial scalar field appear. Even more, the black hole uniqueness for these hairy solutions is violated -- there are regions where more than one black hole solutions with a nontrivial scalar field exist. The global picture is presented in Fig. \ref{fig:phiH(M)} where only the first three scalarized black hole branches are shown to have better visibility. We will call the Schwarzschild solution the trivial branch of solutions (with a trivial scalar field) while the rest of the branches of black holes with a nontrivial scalar field will be called nontrivial branches (with a nontrivial scalar field). The hairy black holes are solutions with a secondary scalar hair since the scalar charge, presented in Fig. \ref{fig:D(M)}, is not independent -- it is a function of the black hole mass. Both Fig. \ref{fig:phiH(M)} and Fig. \ref{fig:D(M)} are symmetric with respect to the x-axis which is a natural consequence of the fact that the coupling function is an even function of $\varphi$, i.e. we have the $Z_2$ symmetry $\varphi \to -\varphi$. Thus for a fixed $M$ the solutions with positive and negative values of $\varphi_H$ would naturally have opposite signs of the scalar charge, but they have the same metric functions and thus mass.

Since the hairy black hole branches are related to the instability of the Schwarzschild solution and bifurcate from it, it is natural to expect that they correspond to the bound states of the potential in the perturbation equation describing the scalar perturbation of the Schwarzschild solution within scalar-Gauss-Bonnet gravity. The different branches of solutions will then have a scalar field with a different number of zeros similar to the eigenfunctions of the perturbation equation \cite{Doneva:2017bvd}. Indeed, the first branch (the red dashed line in Figs. \ref{fig:phiH(M)} and \ref{fig:D(M)}),  called the fundamental branch, is characterized by a scalar field $\varphi(r)$ which has are no zeros (no nodes). The next one (green line) has one zero, the third one (blue line) has two zeros and so on. For smaller values of $M$ there are more bifurcation points but the corresponding non-trivial branches would be even shorter and that is why they are not displayed in the figures. In the case under consideration, these observations points towards the idea that the different nontrivial branches of black hole solutions can be classified by the number of nodes (zeros) of the scalar field.

The linear stability analysis shows that only the fundamental branch of hairy black holes is the stable one while the other branches are linearly unstable. Knowing this fact, it is tempting, at least from a physical point of view, to use the linear stability of black holes as an additional ``parameter'' in the classification problem. Accepting this point of view, we see that in our case, the black hole uniqueness is restored. Unfortunately, this holds only for the coupling function at hand. For other coupling functions (admitting scalarization) the Schwarzschild solution and hairy black holes can coexist for a certain range of the mass, both being linearly stable \cite{Blazquez-Salcedo:2022omw}.  The picture for static non-spherically symmetric and for rotating hairy black holes in scalar-Gauss-Bonnet gravity admitting spontaneous scalarization is similar though a bit more complicated \cite{Collodel:2019kkx}. 

The above example clearly shows that the classification of hairy black holes is a rather challenging problem. The mathematical methods developed so far seem to be insufficient and new mathematical techniques have to be devised.

\medskip
\noindent

\noindent {\bf Acknowledgements:} This study is in part financed by the European Union-NextGenerationEU, through the National Recovery and Resilience Plan of the Republic of Bulgaria, project No. BG-RRP-2.004-0008C01. DD acknowledges financial support via an Emmy Noether Research Group funded by the German Research Foundation (DFG) under grant No. DO 1771/1-1.


\begin{thebibliography}{99}
	
\bibitem{Robinson_2009} D. Robinson, in The Kerr Spacetime: Rotating Black Holes in General Relativity, edited by D. Wiltshire et. al, Cambridge University Press, 2009.	
	
\bibitem{Mazur:2000pn}
P.~O.~Mazur,
arXiv:hep-th/0101012 [hep-th]
	
	
\bibitem{Ruffini:1971bza}
R.~Ruffini and J.~A.~Wheeler,
Phys. Today \textbf{24},  30 (1971)	
	
\bibitem{Israel:1967wq}
W.~Israel,
Phys. Rev. \textbf{164}, 1776-1779 (1967)

\bibitem{Carter:1971zc}
B.~Carter,
Phys. Rev. Lett. \textbf{26}, 331-333 (1971)


\bibitem{Robinson:1975bv}
D.~C.~Robinson,
Phys. Rev. Lett. \textbf{34}, 905-906 (1975)


\bibitem{Mazur:1982db}
P.~O.~Mazur,
J. Phys. A \textbf{15}, 3173-3180 (1982)

\bibitem{Bunting_1983} G.L. Bunting, ``Proof of the uniqueness conjecture for black holes'', PhD Thesis, Univ. of New England,
Armidale, N.S.W., 1983


\bibitem{Heusler_1996} M. Heusler, ``Black hole uniqueness theorems'', Cambridge University Press, 1996


\bibitem{Droz:1991cx}
S.~Droz, M.~Heusler and N.~Straumann,
Phys. Lett. B \textbf{268}, 371-376 (1991)


\bibitem{Volkov:1989fi}
M.~S.~Volkov and D.~V.~Galtsov,
JETP Lett. \textbf{50}, 346-350 (1989)

\bibitem{Bizon:1990sr}
P.~Bizon,
Phys. Rev. Lett. \textbf{64}, 2844-2847 (1990)


\bibitem{Greene:1992fw}
B.~R.~Greene, S.~D.~Mathur and C.~M.~O'Neill,
Phys. Rev. D \textbf{47}, 2242-2259 (1993)

\bibitem{Hartmann:2001ic}
B.~Hartmann, B.~Kleihaus and J.~Kunz,
Phys. Rev. D \textbf{65}, 024027 (2002)

\bibitem{Hollands:2007aj}
S.~Hollands and S.~Yazadjiev,
Commun. Math. Phys. \textbf{283}, 749-768 (2008)



\bibitem{Nedkova:2024oqh}
P.~Nedkova and S.~Yazadjiev,
Lect. Notes Phys. \textbf{1031}, pp. (2024)


\bibitem{Herdeiro:2014goa}
C.~A.~R.~Herdeiro and E.~Radu,
Phys. Rev. Lett. \textbf{112}, 221101 (2014)



\bibitem{Collodel:2020gyp}
L.~G.~Collodel, D.~D.~Doneva and S.~S.~Yazadjiev,
Phys. Rev. D \textbf{102}, 084032 (2020)



\bibitem{Herdeiro:2023roz}
C.~A.~R.~Herdeiro and E.~Radu,
Phys. Rev. Lett. \textbf{131}, 121401 (2023)

\bibitem{Doneva:2022ewd}
D.~D.~Doneva, F.~M.~Ramazano\u{g}lu, H.~O.~Silva, T.~P.~Sotiriou and S.~S.~Yazadjiev,
Rev. Mod. Phys. \textbf{96}, 015004 (2024)



\bibitem{Misner:1973prb}
C.~W.~Misner, K.~S.~Thorne and J.~A.~Wheeler,
``Gravitation,''
W. H. Freeman (1973)

\bibitem{Wald:1984}
R.~Wald,
``General Relativity'',
University of Chicago Press, Chicago (1984).


\bibitem{Kaluza_1921}
T . Kaluza, Sitz. Preuss. Akad. Wiss. Berlin. Math. Phys. S.966 (1921).	

\bibitem{Klein_1926} 
O. Klein, Zeit, f. Phys. Bd.37, S.895 (1926).	


\bibitem{Jordan_1949}
P. Jordan, Nature 164, 637 (1949); Schwerkraft und Weltall (Braunschweig:
Viewrg) (1955)

\bibitem{Fierz:1956zz}
M.~Fierz,
Helv. Phys. Acta \textbf{29}, 128-134 (1956)


\bibitem{Witten:1981me}
E.~Witten,
Nucl. Phys. B \textbf{186}, 412 (1981)



\bibitem{Mallik:1984wc}
S.~Mallik,
Helv. Phys. Acta \textbf{58}, 1004 (1985)
BUTP-84/24-BERN.


\bibitem{Duff:1986hr}
M.~J.~Duff, B.~E.~W.~Nilsson and C.~N.~Pope,
Phys. Rept. \textbf{130}, 1-142 (1986)




\bibitem{Brans:1961sx}
C.~Brans and R.~H.~Dicke,
Phys. Rev. \textbf{124}, 925-935 (1961)

\bibitem{Dicke:1961gz}
R.~H.~Dicke,
Phys. Rev. \textbf{125}, 2163-2167 (1962)




\bibitem{Bergmann:1968ve}
P.~G.~Bergmann,
Int. J. Theor. Phys. \textbf{1}, 25-36 (1968)


\bibitem{Wagoner:1970vr}
R.~V.~Wagoner,
Phys. Rev. D \textbf{1}, 3209-3216 (1970)

\bibitem{Will:2018bme}
C.~M.~Will,
``Theory and Experiment in Gravitational Physics,''
Cambridge University Press (2018)


\bibitem{Wands:1993uu}
D.~Wands,
Class. Quant. Grav. \textbf{11}, 269-280 (1994)



\bibitem{Damour:1992we}
T.~Damour and G.~Esposito-Farese,
Class. Quant. Grav. \textbf{9}, 2093-2176 (1992)



\bibitem{Charmousis:2008kc}
C.~Charmousis,
Lect. Notes Phys. \textbf{769}, 299-346 (2009)


\bibitem{Metsaev:1987zx}
R.~R.~Metsaev and A.~A.~Tseytlin,
Nucl. Phys. B \textbf{293}, 385-419 (1987)



\bibitem{Yunes:2011we}
N.~Yunes and L.~C.~Stein,
Phys. Rev. D \textbf{83}, 104002 (2011)


\bibitem{Pani:2011gy}
P.~Pani, C.~F.~B.~Macedo, L.~C.~B.~Crispino and V.~Cardoso,
Phys. Rev. D \textbf{84}, 087501 (2011)



\bibitem{Alexander:2009tp}
S.~Alexander and N.~Yunes,
Phys. Rept. \textbf{480}, 1-55 (2009)
[arXiv:0907.2562 [hep-th]].



\bibitem{Horndeski:1974wa}
G.~W.~Horndeski,
Int. J. Theor. Phys. \textbf{10}, 363-384 (1974)


\bibitem{Deffayet:2009mn}
C.~Deffayet, S.~Deser and G.~Esposito-Farese,
Phys. Rev. D \textbf{80}, 064015 (2009)


\bibitem{Kobayashi:2011nu}
T.~Kobayashi, M.~Yamaguchi and J.~Yokoyama,
Prog. Theor. Phys. \textbf{126}, 511-529 (2011)




\bibitem{Armendariz-Picon:1999hyi}
C.~Armendariz-Picon, T.~Damour and V.~F.~Mukhanov,
Phys. Lett. B \textbf{458}, 209-218 (1999)

\bibitem{Chiba:1999ka}
T.~Chiba, T.~Okabe and M.~Yamaguchi,
Phys. Rev. D \textbf{62}, 023511 (2000)


\bibitem{Armendariz-Picon:2000nqq}
C.~Armendariz-Picon, V.~F.~Mukhanov and P.~J.~Steinhardt,
Phys. Rev. Lett. \textbf{85}, 4438-4441 (2000)





\bibitem{Armendariz-Picon:2005oog}
C.~Armendariz-Picon and E.~A.~Lim,
JCAP \textbf{08}, 007 (2005)



\bibitem{Rendall:2005fv}
A.~D.~Rendall,
``Dynamics of k-essence,''
Class. Quant. Grav. \textbf{23}, 1557-1570 (2006)

\bibitem{Bernard:2019fjb}
L.~Bernard, L.~Lehner and R.~Luna,
Phys. Rev. D \textbf{100}, 024011 (2019)



\bibitem{Galloway:1999bp}
G.~J.~Galloway, K.~Schleich, D.~M.~Witt and E.~Woolgar,
Phys. Rev. D \textbf{60}, 104039 (1999)


\bibitem{Hawking:1973uf}
S.~W.~Hawking and G.~F.~R.~Ellis,
``The Large Scale Structure of Space-Time,''
Cambridge University Press (2023)


\bibitem{Chrusciel:1996bj}
P.~T.~Chrusciel,
Commun. Math. Phys. \textbf{189}, 1-7 (1997)

\bibitem{Friedrich:1998wq}
H.~Friedrich, I.~Racz and R.~M.~Wald,
Commun. Math. Phys. \textbf{204}, 691-707 (1999)

\bibitem{Bunting_1987}	
G. Bunting and A. Masood-ul Alam, Gen. Rel. Grav. 19, No2, 147 (1987).


\bibitem{Schon:1979rg}
R.~Schoen and S.~T.~Yau,
Commun. Math. Phys. \textbf{65}, 45-76 (1979)

\bibitem{Witten:1981mf}
E.~Witten,
Commun. Math. Phys. \textbf{80}, 381 (1981)	


\bibitem{Hollands:2008fm}
S.~Hollands and S.~Yazadjiev,
Commun. Math. Phys. \textbf{302}, 631-674 (2011)


\bibitem{Mazur:1984}
P. Mazur,
``A Global Identity for Nonlinear Sigma-Models,''
Phys. Lett A 100 (1984) 341.


\bibitem{Weinstein:1995tg}
G.~Weinstein,
Math. Res. Lett. \textbf{3}, 835-844 (1996)

\bibitem{Hawking:1972qk}
S.~W.~Hawking,
Commun. Math. Phys. \textbf{25}, 167-171 (1972)	


\bibitem{Bekenstein:1972ny}
J.~D.~Bekenstein,
Phys. Rev. Lett. \textbf{28}, 452-455 (1972)


\bibitem{Sotiriou:2011dz}
T.~P.~Sotiriou and V.~Faraoni,
Phys. Rev. Lett. \textbf{108}, 081103 (2012)


\bibitem{Herdeiro:2015waa}
C.~A.~R.~Herdeiro and E.~Radu,
Int. J. Mod. Phys. D \textbf{24}, 1542014 (2015)


\bibitem{Heusler:1992ss}
M.~Heusler,
J. Math. Phys. \textbf{33}, 3497-3502 (1992)

\bibitem{Bekenstein:1995un}
J.~D.~Bekenstein,
Phys. Rev. D \textbf{51},  R6608 (1995)

\bibitem{Sudarsky:1995zg}
D.~Sudarsky,
Class. Quant. Grav. \textbf{12}, 579-584 (1995)






\bibitem{Cardoso:2013fwa}
V.~Cardoso, I.~P.~Carucci, P.~Pani and T.~P.~Sotiriou,
Phys. Rev. Lett. \textbf{111}, 111101 (2013)




\bibitem{Hod:2012px}
S.~Hod,
Phys. Rev. D \textbf{86}, 104026 (2012)
[erratum: Phys. Rev. D \textbf{86}, 129902 (2012)]



\bibitem{Herdeiro:2015gia}
C.~Herdeiro and E.~Radu,
Class. Quant. Grav. \textbf{32}, 144001 (2015)


\bibitem{Heusler:1993cd}
M.~Heusler,
Class. Quant. Grav. \textbf{10}, 791-799 (1993)





\bibitem{Doneva:2020dji}
D.~Doneva and S.~S.~Yazadjiev,
Phys. Rev. D \textbf{102}, 084055 (2020)




\bibitem{Yazadjiev:2024rql}
S.~S.~Yazadjiev and D.~D.~Doneva,
Eur. Phys. J. C \textbf{84}, 492 (2024)



\bibitem{Pena:1997cy}
I.~Pena and D.~Sudarsky,
Class. Quant. Grav. \textbf{14}, 3131-3134 (1997)




\bibitem{Friedberg:1976me}
R.~Friedberg, T.~D.~Lee and A.~Sirlin,
Phys. Rev. D \textbf{13}, 2739-2761 (1976)

\bibitem{Jetzer:1991jr}
P.~Jetzer,
Phys. Rept. \textbf{220}, 163-227 (1992)	

\bibitem{Liebling:2012fv}
S.~L.~Liebling and C.~Palenzuela,
Living Rev. Rel. \textbf{26},  1 (2023)	

\bibitem{Collodel:2019uns}
L.~G.~Collodel, D.~D.~Doneva and S.~S.~Yazadjiev,
Phys. Rev. D \textbf{101},  044021 (2020)	


\bibitem{Heusler:1995qj}
M.~Heusler,
Class. Quant. Grav. \textbf{12}, 2021-2036 (1995)



\bibitem{Graham:2014mda}
A.~A.~H.~Graham and R.~Jha,
Phys. Rev. D \textbf{89},  084056 (2014)
[erratum: Phys. Rev. D \textbf{92},  069901 (2015)]



\bibitem{Graham:2014ina}
A.~A.~H.~Graham and R.~Jha,
Phys. Rev. D \textbf{90},  041501 (2014)

\bibitem{Hui:2012qt}
L.~Hui and A.~Nicolis,
Phys. Rev. Lett. \textbf{110}, 241104 (2013)


\bibitem{Sotiriou:2013qea}
T.~P.~Sotiriou and S.~Y.~Zhou,
Phys. Rev. Lett. \textbf{112}, 251102 (2014)


\bibitem{Capuano:2023yyh}
L.~Capuano, L.~Santoni and E.~Barausse,
Phys. Rev. D \textbf{108}, 6 (2023)



\bibitem{Babichev:2013cya}
E.~Babichev and C.~Charmousis,
JHEP \textbf{08}, 106 (2014)



\bibitem{Doneva:2017bvd}
D.~D.~Doneva and S.~S.~Yazadjiev,
Phys. Rev. Lett. \textbf{120},  131103 (2018)




\bibitem{Blazquez-Salcedo:2022omw}
J.~L.~Bl\'azquez-Salcedo, D.~D.~Doneva, J.~Kunz and S.~S.~Yazadjiev,
Phys. Rev. D \textbf{105}, 124005 (2022)


\bibitem{Collodel:2019kkx}
L.~G.~Collodel, B.~Kleihaus, J.~Kunz and E.~Berti,
Class. Quant. Grav. \textbf{37}, 075018 (2020)









\end{thebibliography}
\end{document}